\documentclass[letterpaper,oldversion]{aa}  
\usepackage{graphicx}
\usepackage{txfonts}

\def\kms{{km~s$^{-1}$}}

\def\degr{\hbox{$^\circ$}}
\def\arcmin{\hbox{$^\prime$}}
\def\arcsec{\hbox{$^{\prime\prime}$}}
\def\utw{\smash{\rlap{\lower5pt\hbox{$\sim$}}}}
\def\udtw{\smash{\rlap{\lower6pt\hbox{$\approx$}}}}

\def\fdg{\hbox{$.\!\!^\circ$}}

\begin{document}
   \title{The Westerbork SINGS Survey}

   \subtitle{I. Overview and Image Atlas}

   \author{R. Braun
          \inst{1}
          \and
	  T.A. Oosterloo
	  \inst{1}
	  \and
	  R. Morganti
	  \inst{1}
	  \and
	  U. Klein
	  \inst{2}
	  \and
	  R. Beck
	  \inst{3}
          }

   \offprints{R. Braun}

   \institute{ASTRON, PO Box 2, 7990 AA Dwingeloo, The Netherlands
	      \and
	      Argelander-Institut f\"ur Astronomie, Radio Astronomy
	      Department, Auf dem H\"ugel 71, D-53121 Bonn, Germany
              \and
              Max-Planck-Institut f\"ur Radioastronomie, Auf dem
	      H\"ugel 69, 53121 Bonn, Germany
             }

   \date{Received ; accepted }

 
  \abstract {We have obtained moderately deep radio continuum imaging
   at 18 and 22cm with the Westerbork array of 34 nearby galaxies
   drawn from the Spitzer SINGS and Starburst samples to enable
   complimentary analysis. The sub-sample have an optical major axis
   diameter in excess of 5 arcmin and are North of Declination
   12\fdg5. Sub-sample galaxies span a very wide range of
   morphological types and star formation rates. Resolved detection
   was possible for every galaxy. This constitutes a first time
   detection at GHz radio frequencies for about half of the
   sample. Analysis of both total intensity and polarization
   properties of the sample will be published in companion papers.
   Both the \ion{H}{i} and OH main-lines of the target galaxies were
   within the observed band-pass, albeit with only coarse velocity
   resolution. Only two low mass elliptical galaxies were undetected
   in \ion{H}{i}. Four of the sub-sample galaxies were detected in OH
   main-line absorption, including two new detections. The results are
   presented in the form of an image atlas for which a standard
   transfer function and image size are used throughout and whereby
   the radio continuum, DSS optical and integrated \ion{H}{i} are
   displayed side-by-side. Continuum and \ion{H}{i} line photometry
   are tabulated for all targets. } 


   \maketitle
%

\section{Introduction}

An understanding of the star formation process is critical to a wide
range of astrophysical questions, from the phase and energy balance in
the local ISM, to the circumstances surrounding the universal peak of
star formation and merger activity that apparently typify the early
universe. The traditional global tracers of current star formation in
galaxies, like H$\alpha$ imaging, have led to some empirical
descriptions of how SF occurs (eg. Kennicutt \cite{kenn98}).  However,
given the complexity of the many competing processes ocurring in a
galactic disk, it is clear that a coordinated effort, including
multiple ISM tracers, will be needed to make more substantial
progress. Just such an initiative is currently underway with the {\it
Spitzer\ } Nearby Galaxy Survey (SINGS, Kennicutt et
al. (\cite{kenn03})). This {\it Spitzer\ } Legacy program provides
pixel-resolved SED data from the visible to 160 $\mu$m for a sample of
75 galaxies nearer than 30~Mpc that span the largest possible range of
Hubble types and (to a lesser extent) SFR's. In addition to the {\it
Spitzer\ } data, the SINGS program will make a comprehensive set of
broad-band optical BVRIJHK, narrow-band H$\alpha$, as well as UV, CO
and (in some cases) HI data available to the general public.

An important supplement to the SINGS program is sensitive radio
continuum imaging of the sample galaxies. In this paper we describe
the acquisition of such data with the Westerbork Synthesis Radio
Telescope (WSRT) for a sub-sample of thirty SINGS galaxies,
supplemented with a sub-sample of five Starburst galaxies. Radio
continuum data has been acquired in the 20cm band which reaches a
3$\sigma$ Emission Measure (EM) sensitivity of 75~pc~cm$^{-6}$ for
emission filling the 1~kpc beam (at the average galaxy distance of
10~Mpc). For comparison, the Diffuse Ionized Gas which fills the disks
of spiral galaxies has a typical peak EM determined from H$\alpha$
imaging of about 100~pc~cm$^{-6}$, while discrete HII region complexes
reach EM's of 10$^3-10^4$~pc~cm$^{-6}$ (eg. Greenawalt et
al. \cite{gree98}). Our continuum sensitivity is thus sufficient to
detect both diffuse and discrete regions of thermal emission within
our target galaxy disks in addition to the much brighter non-thermal
component which typically accounts for about 75\% of the total galaxy
flux at these frequencies.

Amongst the various scientific goals which will be addressed with our
newly acquired data are a resolved study of the FIR-radio correlation
(for which the first analysis paper has just appeared in Murphy et
al. \cite{murp06}), a systematic study of galaxy magnetic fields and
an unbiased survey of HI and OH emission in both the target galaxies as
well as their extended environment. These and other topics will
addressed in forthcoming papers.

The current paper is organized as follows. We describe the sample
definition, observations and reduction in \S\ref{sec:obse} and present
the results in \S\ref{sec:resu}. Digital images will be made available
from the {\it Spitzer\ } SINGS web site at
http://data.spitzer.caltech.edu/popular/sings/.


\section{Observations and Reduction}
\label{sec:obse}

The SINGS sample (Kennicutt et al. \cite{kenn03}) includes 75
``normal'' galaxies within 30~Mpc that span (as uniformly as possible)
Hubble types from Irregular to Elliptical and star formation rates
from less than 0.001 to more than 10 M$_\odot$yr$^{-1}$.  We define
our sub-sample of SINGS galaxies by choosing those North of
Declination 12\fdg5 and for which D$_{25}$ (the optical B band
isophotal diameter at a brightness of 25 mag per square arcsec) is
greater than 5~arcmin. These criteria were used to insure that the
angular resolution across the targets would be sufficient to enable
resolved study, given the limitations of the WSRT array (an east-west
configuration of 2.7~km extent). The resulting thirty SINGS galaxies
were supplemented by applying the same selection criteria to the {\it
Spitzer\ } Starburst GTO program of G. Rieke
(http://ssc.spitzer.caltech.edu/geninfo/gto/). This yielded four
additional targets which extend our sample into the very high star
formation rate regime.

Each target was observed for one (and occasionally two) twelve hour
integration(s) in the "maxi-short" array configuration, in which
shortest East-West baselines of 36, 54, 72 and 90 meters are all
measured simultaneously as well as a longest baseline of about 2700
m.~Details of observing sessions can be found in
Table~\ref{tab:obse}. The target observations were bracketed by
observations of the primary total intensity and polarization
calibration sources 3C147 and 3C286, as well as additional calibrators
CTD93 and 3C138, yielding an absolute flux density calibration
accuracy of better than 5\%. The observing frequency was switched
every five minutes between two settings (1300 to 1432 MHz and 1631 to
1763 MHz). Each frequency setting was covered with eight sub-bands of
20~MHz nominal width, but spaced by 16~MHz to provide contiguous,
non-attenuated coverage. An effective integration time of 6 (or 12)
hours was realized at each frequency setting. All four polarizations
products and 64 spectral channels were obtained in each
sub-band. After careful editing of incidental radio frequency
interference (RFI), external total intensity, band-pass and
polarization calibration of the data was performed in the AIPS
package.

Subsequently, each field was self-calibrated in Stokes I, Q and U using
an imaging pipeline based on the Miriad package. Each of the eight
sub-bands for a given frequency setting was first processed and imaged
independently; and these were subsequently combined with an inverse
variance weighting. Deconvolution of each sub-band image was performed
iteratively within a threshold mask based on a spatial smoothing of
the previous iterate. The individual frequency channels (of 312.5 kHz
width) were gridded during imaging, so that band-width smearing
effects were negligible. Before combination, each deconvolved sub-band
image was subjected to a linear deconvolution of the Gaussian
restoring beam followed by Gaussian tapering to a nominal PSF size as
well as a primary beam correction appropriate to the sub-band
observing frequency. The final combined image Gaussian PSF parameters
and actual {\sc RMS} noise levels are shown in Table~\ref{tab:obse}.

A different reduction strategy was employed for NGC~3031. The diffuse
emission from this galaxy was so extended (filling the entire
telescope primary beam) that the individual sub-band images were
strongly self-confused. This severely hampered self-calibration and
imaging of the individual sub-bands. The best results for this galaxy
were obtained by including the data from all 16 sub-bands (both 18~cm
and 22~cm) simultaneously in the the self-calibration and imaging
process (using the Miriad task mfclean). Consequently, only a single
final image (at effective frequency 1589 MHz) is available for this
galaxy.

For the bright starburst galaxy NGC~3034, only the 18~cm band data
could be successfully calibrated and imaged. 

As can be seen from Table~\ref{tab:obse}, the noise performance is
significantly better for most targets, by about 50\%, in the 22~cm
band than in the 18~cm band. This is partly due to a higher system
temperature of the receiving system near 1700~MHz than near 1400~MHz,
but is also due to higher levels of RFI near 1700~MHz. Residual RFI
and calibration imperfections account for the variations in final
image quality. A few particularly bad examples are the 18~cm images of
NGC~4125 and NGC~4826. The noise performance in the 22~cm band is in
all cases somewhat worse than the theoretical value of 15~$\mu$Jy (for
an ideal 6 hour integration) partly due to the data weighting strategy
employed (a semi-uniform weighting using a robustness parameter of
$-$1 in the Miriad task Invert), but also due to source and sidelobe
confusion effects in these complex fields. Repeat observations were
obtained for those targets which had substantial data loss in the
original coverage due to RFI or equipment failures.

The two observing bands were chosen to provide frequency coverage of
the neutral hydrogen (at 22~cm) and hydroxyl (at 18~cm) emission lines
at the recession velocities of all targets as well as their extended
environment. The 312.5 kHz frequency resolution, corresponding to
about 66~km~s$^{-1}$ in the \ion{H}{i} line, is rather coarse, while
the total velocity coverage of some 27000~km~s$^{-1}$, is extremely
broad.  The expected velocity range of the \ion{H}{i} line due to our
own Galaxy and that of each target galaxy was excluded from the
continuum imaging described above. An effective continuum model was
available for each observed sub-band (as a by-product of the
self-calibration step described above). The continuum model was
subtracted from the visibility data of each sub-band. The residual
data were imaged, hanning smoothed in velocity, deconvolved and
subjected to a residual continuum subtraction in the image plane. The
resulting line image cubes have the PSF and {\sc RMS} properties
listed in Table~\ref{tab:obse}. The PSF dimensions are those that
result from a semi-natural data weighting (a robustness parameter of
+0.5 in the Miriad task Invert). Images of integrated \ion{H}{i}
emission were generated by summing masked channel images at nominal
angular resolution. The mask for each channel was generated by forming
a convolved channel image with 120~arsec resolution and including in
the mask those regions which exceeded five times the {\sc RMS} level
in either the smoothed or the original image. This is a very
conservative limit on \ion{H}{i} detection; normally a threshold in
smoothed images of between 2 and 3 $\sigma$ is employed. However, in
view of the very coarse intrinsic velocity resolution and semi-natural
nominal PSF this criterion was found to be quite effective in
practise and eliminated the need for any interactive image
blanking. Manual inspection of the channel images, particularly for
the two undetected targets did not reveal any significant emission
features missed by our masking criterion.

\section{Results}
\label{sec:resu}

An overview of the imaging results is given in
Figs.~\ref{fig:fourset1} -- \ref{fig:twoset}, where radio continuum,
optical R band and integrated \ion{H}{i} emission are displayed for
our sample galaxies. The same transfer function (as indicated in the
Figure captions) and standard image size (28$\times$28 arcmin) have
been used for all galaxies to facilitate comparison. The median was
subtracted from each optical DSS image before display. The only minor
complication to direct inter-comparison is the choice of a fixed scale
in flux units (Jy/Beam) rather than surface brightness units in the
presentation of the radio continuum images since the beam area
varies by almost a factor of five between low Declination and high
Declination targets. We comment briefly on each of the targets in
\S\ref{subsec:gals} below.

\begin{figure*}
   \centering
   \includegraphics[width=5.6cm]{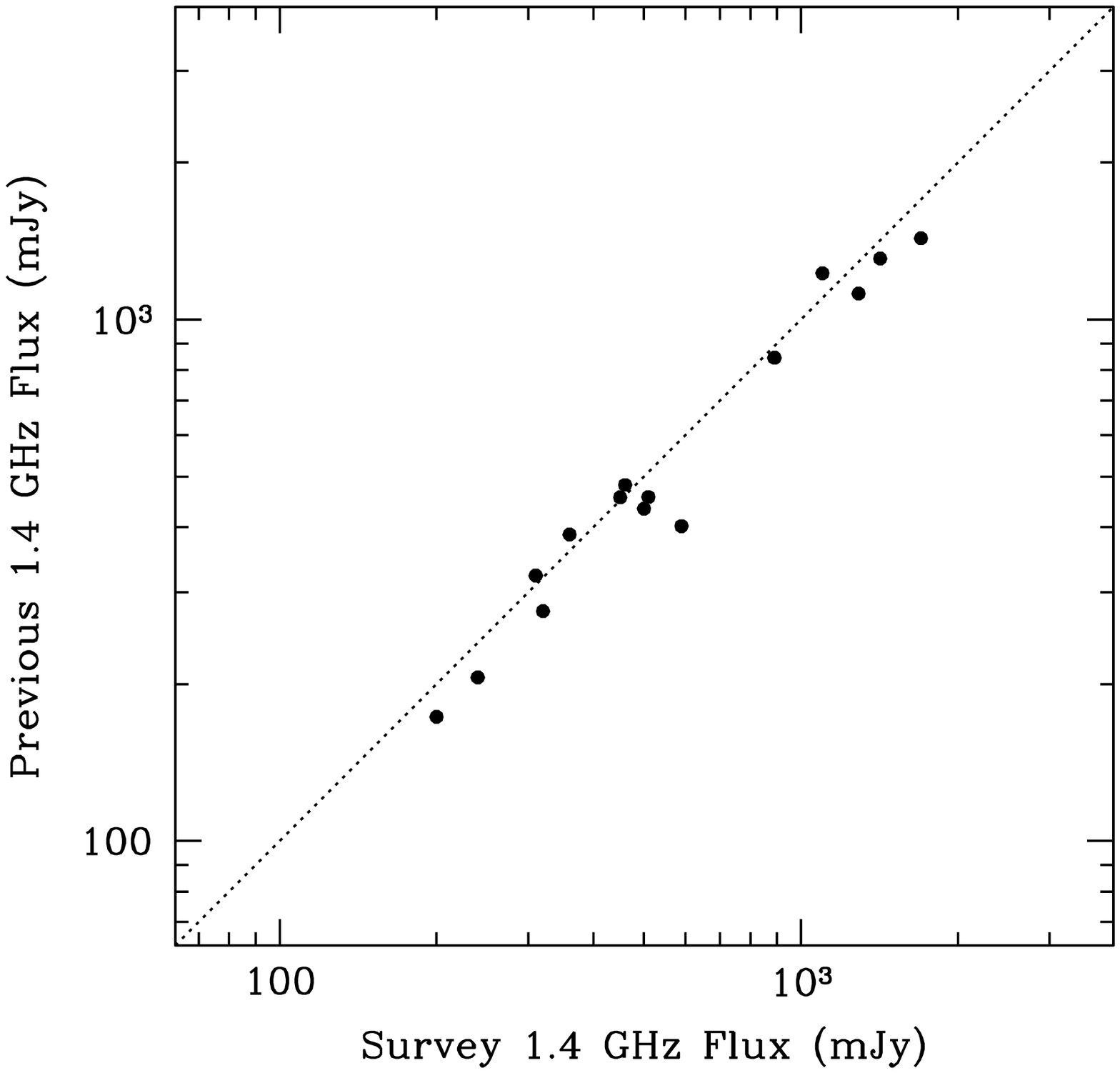} 
   \includegraphics[width=5.6cm]{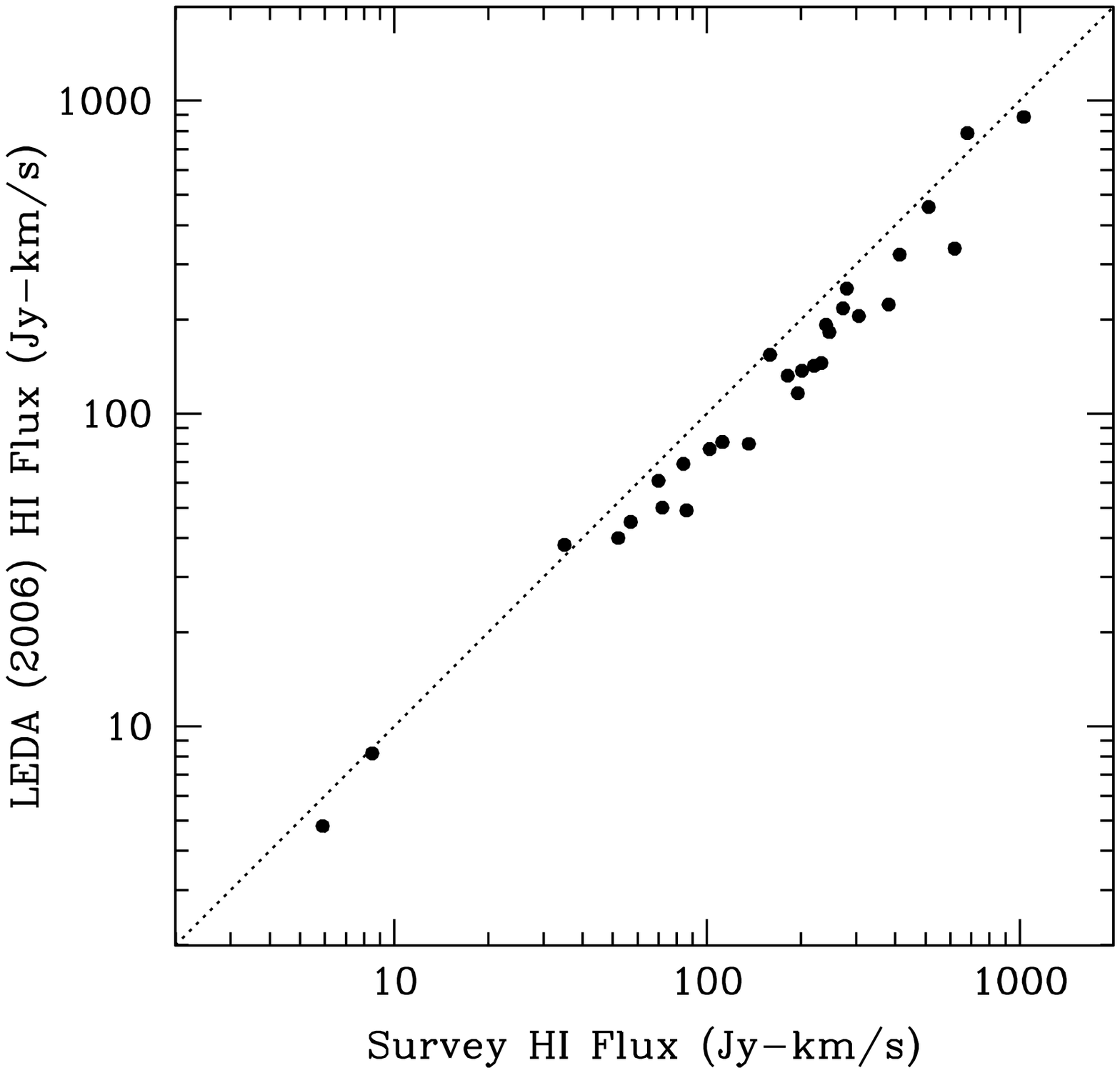}
   \includegraphics[width=5.6cm]{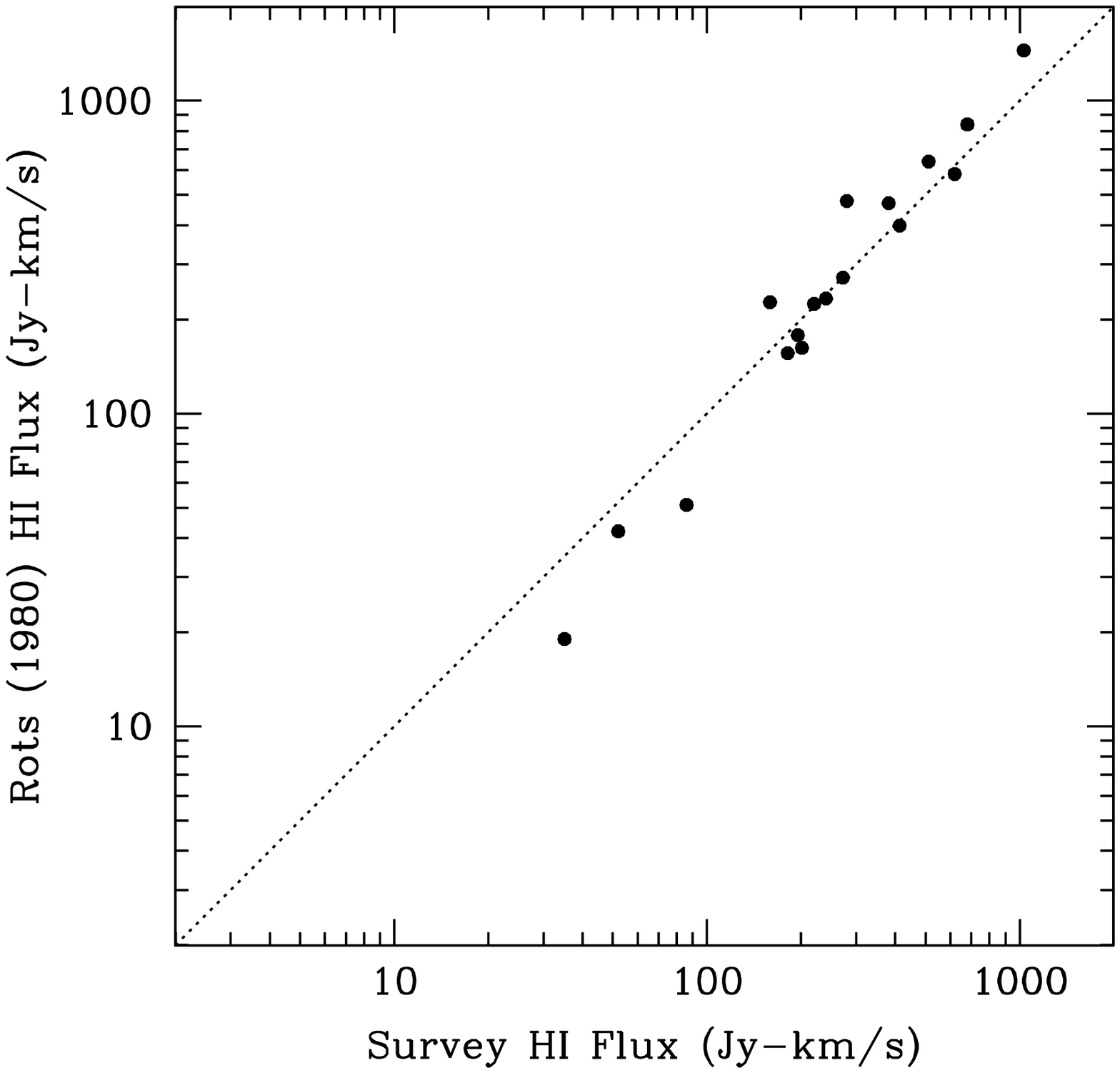}
      \caption{Radio continuum (left), and neutral hydrogen (center
              and right) flux comparison of the current survey with
              previously published values. The comparison data is that
              of White \& Becker (\cite{whit92}) for the radio
              continuum and LEDA (http:/leda.univ-lyon1.fr) (center)
              and Rots (\cite{rots80}) (right) for the neutral hydrogen. }
         \label{fig:flxcmp}
   \end{figure*}

The total detected flux density in the 22~cm continuum band has been
determined for each target galaxy and is tabulated in
Table~\ref{tab:prop}, excluding NGC~3031 and NGC~3034, for which
appropriate 22~cm images were not available. The aperture radius and
flux integration method are indicated in the Table for each galaxy. In
those cases where no extended disk emission was detected, a simple
rectangular ``Box'' or ``Poly''-gon was integrated with the noted
effective radius. In a single case (NGC~4125) an elliptical Gaussian
was fit to the detected source; method ``GFit''. In most cases, the
diffuse brightness distribution was integrated in concentric circles,
whereby a background brightness level was estimated and subtracted
from the integral; method ``Rad''. The background estimate was
determined at radii extending 10\% beyond the radius indicated in the
Table. In those cases where particularly bright background radio
sources were present in the field, such sources were first
interactively blanked from the image, before integration in concentric
circles as above; method ``Blk/Rad''. The error estimates in the table
reflect the variation in total flux that result from alternate choices
of the integation radius.  Although all of these values either agree
with or slightly exceed (by up to 20\%) current estimates in the
literature (White \& Becker \cite{whit92}) they must still be regarded
as lower limits, since the brightness distribution declines so
smoothly into the noise floor. Our measured fluxes are compared with
those of White \& Becker in Fig.~\ref{fig:flxcmp}. About half of the
entries in our Table correspond to first time detections of continuum
emission from these galaxies at GHz radio frequencies.

The integrated \ion{H}{i} line flux is also listed in
Table~\ref{tab:prop} for each galaxy together with the peak observed
column density. The error estimates in total \ion{H}{i} line flux
include uncertainties due to our conservative masking criterion. In
view of the very low recession velocites of NGC~3031 and NGC~3034, the
\ion{H}{i} emission from these galaxies was strongly confused with
that of our Galaxy and is not shown or tabulated. The only
non-detections in \ion{H}{i} are NGC~4125 and NGC~4552 with the
indicated 3$\sigma$ upper limits. Our measured \ion{H}{i} fluxes
are compared with those in the literature in Fig.~\ref{fig:flxcmp},
where we have taken the LEDA database entries
(http:/leda.univ-lyon1.fr) from June 2006 and those of Rots
(\cite{rots80}) for comparison. As can be seen in the figure, our
fluxes either agree with, or exceed (by as much as a factor of 1.8)
those tabulated in LEDA. Comparison with the Rots values indicates
that only the two galaxies of largest angular size in our sample
(NGC~2403 and 5055) have significantly low detected fluxes (by factors
of 1.4 and 1.7). There is also a discrepancy of a factor of 1.4 in the
NGC~5194 field, due to the extended tidal debris and companions in
this region. Smaller underestimates (by a factor of 1.2) are present
in the cases of NGC~628 and 6946. We recognize that our flux values
must be considered as lower limits, given our finite sensitivity and
incomplete visibility sampling. We have marked the NGC~2403 and 5055
values with a ``:'' suffix in Table~\ref{tab:prop} to indicate the
poor quality of these estimates. The effective velocity resolution in
the \ion{H}{i} line cubes is 132~km~s$^{-1}$, which is so coarse that
significant sensitivity loss will be experienced for intrinsically
narrow lines (such as the 25~km~s$^{-1}$ {\sc FWHM} of a thermal
10$^4$~K gas).

\subsection{Notes on Individual Galaxies}
\label{subsec:gals}

{\it Holmberg II\ } Diffuse, low-level radio continuum emission is
detected in the region of highest stellar surface density of this
galaxy as well as several higher brightness features to the East.  The
work of Tongue \& Westphal (\cite{tong95}) shows that all of these
brighter features appear to be associated with H$\alpha$-emitting
regions; the fainter ones are likely to be purely thermal, while the
brighter ones emit a mixture of free-free (from \ion{H}{II} regions)
and synchrotron (from supernovae or remnants) radiation. A classical
double background radio galaxy of FR~II type with an indication of
precessing beams is seen in the South-East.

\noindent{\it IC 2574\ } Diffuse radio continuum emission from the
stellar disk is not detected convincingly. Brighter knots in the
North-East are associated with regions of massive star formation in
the galaxy. They embrace the small, but pronounced, \ion{H}{i} hole in
this galaxy adjoining the region of highest column density. This it is
one of the few documented cases of an \ion{H}{i} hole associated with
a classical hot (X-ray emitting) bubble (Walter et al. \cite{walt98}).

\noindent{\it NGC 628\ } This galaxy possesses a very extended
\ion{H}{i} envelope, most likely fed by infalling intergalactic gas
(Kamphius \& Briggs \cite{kamp92}).  The radio continuum map shows
diffuse disk emission as well as the two main spiral arms at higher
surface brightness. Curiously, the optically bright central region is
deficient in the radio continuum, although the CO map of Wakker \&
Adler (\cite{wakk95}) suggests the presence of copious molecular gas
within the central 30\arcsec\ region, albeit without any concentration
towards the very center.

\noindent{\it NGC 925\ } A low surface brightness disk is detected 
in this asymmetric late-type galaxy. The overall asymmetry may be 
attributed to infall and frequent gravitational encounters of the 
past few gigayears (Pisano et al. \cite{pisa98}). Yet the star 
formation density reflected by the radio continuum brightness is low. 

\noindent{\it NGC 2146\ } Residual calibration errors are responsible
for the ``spoke-like'' artifacts radiating from the nuclear region.
This galaxy is a prototypical merger, evidenced by tidal \ion{H}{i}
structures and a warped optical disk. The streamer seen in the neutral
hydrogen towards the South continues into a very extended tidal tail
(Taramopoulos et al. \cite{tara01}). The merger has resulted in
vigorous star formation throughout this galaxy, resulting in a
brightness comparable to that of M~82. Similar to this nearby
prototype, the bright radio continuum contains numerous compact
sources reflecting ultra-compact \ion{H}{II} regions and/or radio
supernovae (Tarchi et al. \cite{tarc00}). They are also coincident
with the prominent absorption feature seen in the \ion{H}{i} map. The
OH main lines at 1667 and 1665~MHz are detected in absorption against
the nuclear region as shown in Fig.~\ref{fig:hiohabs}. Faint 1667 MHz
absorption may be detected at velocities blue-shifted by up to 200
\kms\  relative to the main disk. The peak line depths at 1667 and 1665
MHz are -3.13 and -1.70 mJy/Beam relative to a continuum brightness of
387 mJy/Beam measured with the same PSF. This 1667/1665 line ratio
correponds closely to the value of 1.8 expected from an optically thin
thermal gas (Heiles \cite{heil69}). The \ion{H}{i} absorption line
depth is -40.3 mJy/Beam relative to the 480 mJy/Beam continuum
brightness. The 10:1 observed ratio of \ion{H}{i} to OH opacity is
typical of that seen in the Galactic survey of Liszt \& Lucas
\cite{lisz96}.

\noindent{\it NGC 2403\ } This late-type spiral is a twin of M~33.
The faint radio continuum from the disk has a number of brighter spots
superimposed, which are coincident with prominent \ion{H}{II} regions
(Drissen et al. \cite{dris99}). The galaxy has become known for the
well-documented ``anomalous'' velocity gas component associated with
the halo (Faternali et al. \cite{frat01}, \cite{frat04}) which may be
the signature of accreting gas.

\noindent{\it NGC 2841\ } A nuclear emission component is detected as
well as a more diffuse hour-glass structure of 5\arcmin\ major-axis
diameter. The two brighter regions mark the apparent waist of this
diffuse structure (see also Fig.~1 in Hummel \& Bosma
\cite{humm82}). This feature is reminiscent of a large diameter
bi-polar outflow along the minor axis, but may well have some other
origin. The \ion{H}{i} map exhibits low column densities within the
stellar disk, and extends well beyond this, but with a different
position angle, indicative of a warp, as has been previously
documented (eg. Bosma \cite{bosm81}).

\noindent{\it NGC 2903\ } This is a grand-design spiral with a bright
central region characterised by a starburst. A high-resolution study 
of the central region was performed by Saikia et al. \cite{saik94}). 
The radio continuum, H$\alpha$ and CO emission are almost identically 
distributed along a central bar (Jackson et al. \cite{jack91}). The 
\ion{H}{i} traces both the inner and outer spiral structures, but 
also extends far beyond the stellar disk, though with a different 
position angle. As in the above case, this is indicative of a 
warp (eg. Wevers et al. \cite{weve86}). 

\noindent{\it NGC 2976\ } Diffuse disk emission is seen in this dwarf
spiral, together with brighter knots at either end of the major axis.
These are coincident with strong H$\alpha$ emission (see the images in
Simon et al. \cite{simo03}). This gives the same overall appearance in
the radio continuum as seen in case of NGC~2841 and may also represent
a similar phenomenon.

\noindent{\it NGC 3031\ } This is the well known galaxy M~81, the dominant
galaxy of the M~81 Group, and it is one of the best-studied
spiral galaxies in all spectral regimes. Compared to other
grand-design spirals, its radio continuum brightness is rather 
low. The overall spiral structure is readily recognisable in the radio
continuum. The low radio continuum brightness and corresponding low
star formation activity is in accord with the low CO surface brightness 
throughout M~81 (Brouillet et al. \cite{brou91}; \cite{brou98}). The
central point source is a low-luminosity AGN (e.g. de Bruyn
\cite{bruy76}, Filippenko \cite{fili88}).

\noindent{\it NGC 3034\ } As M~81's nearest companion, this
prototypical dwarf starburst galaxy has seen countless studies,
from radio wavelengths through $gamma$-ray energies. Similar to
NGC~2146 (see above), although on a smaller scale, its violent
star formation activity is caused by tidal interaction (in this case
with M~81) and the resulting gas infall. Our radio continuum image
delineates the central starburst, with some possible filaments
emerging from the disturbed disk (although spoke-like calibration
artifacts are also present), which were previously reported by Reuter
et al. (\cite{reut92}). The double source to the South-East is a
background radio galaxy.

\noindent{\it NGC 3079\ } Well known for its central activity, this
galaxy produces a bi-conical wind along the minor axis, although of
limited angular extent. The corresponding ``figure-eight'' structure
seen in the radio continuum on smaller scales (Duric \& Seaquist
\cite{duri88}) is not resolved in our 22-cm image, but is apparent as
a protusion East of the nucleus. The \ion{H}{i} map exhibits strong
central absorption and two dwarf galaxies towards the West and
North-West, which were previously reported by Irwin et
al. (\cite{irwi87}). Away from the plane, there are plumes of radio
continuum emission in the North-East and in the South-West, while in
the \ion{H}{i} line a linear structure is apparent, which was not
visible in the lower-resolution VLA maps of Irwin et al. This feature
is suggestive of tidal debris of a recent merger (cf. NGC~4631 below).
OH mainline absorption was detected previously in this galaxy by
Haschick \& Baan (\cite{hasc85}). This is also seen prominently in
Fig.~\ref{fig:hiohabs}. Peak line depths at 1667 and 1665 MHz are
-28.5 and -22.4 mJy/Beam relative to a continuum brightness of 398
mJy/Beam measured with the same PSF. The 1667/1665 line ratio of 1.27
approaches the value of unity expected for an optically thick thermal
gas (Heiles \cite{heil69}). The \ion{H}{i} absorption line depth is
-35.7 mJy/Beam relative to the 486 mJy/Beam continuum brightness. The
1:1 observed ratio of \ion{H}{i} to OH opacity is more extreme in
molecular content than any seen in the Galactic survey of Liszt \&
Lucas \cite{lisz96}.

\noindent{\it NGC 3184\ } Nuclear and faint disk emission are detected 
in this spiral galaxy, within the H$\alpha$-emitting area (see image in
Doane et al. \cite{doan04}). 

\noindent{\it NGC 3198\ } Nuclear and faint disk emission are detected in 
this spiral galaxy, known for the seminal study of its distribution of dark 
matter (van Albada et al. \cite{alba85}; Begeman \cite{bege89}). The 
galaxy is surrounded by an extensive asymmetric \ion{H}{i} envelope. 

\noindent{\it NGC 3627\ } This radio-bright galaxy belongs to the Leo
triplet, together with NGC~3628 (see below) and NGC~3623 (see Zhang et
al.  \cite{zhan93} for an overview). The bright radio continuum is
co-extensive with the bright stellar disk as well as intense H$\alpha$
(Chemin et al. \cite{chem03}), CO (Regan et al. \cite{rega99}), and
FIR (Smith et al. \cite{smit94}) emission. A multi-frequency radio
continuum and polarisation study was presented by Soida et
al. (\cite{soid01}). Weak OH mainline absorption is detected against
the brightest central radio continuum feature (offset by some 20
arcsec to the NE of the nuclear position) an shown in
Fig.~\ref{fig:hiohabs}. Peak line depths at 1667 and 1665 MHz are
-1.46 and -0.93 mJy/Beam relative to a continuum brightness of 31.6
mJy/Beam measured with the same PSF. The 1667/1665 line ratio of 1.6
is intermediate between the optically thin and optically thick cases.
No corresponding \ion{H}{i} absorption is detected.

\noindent{\it NGC 3628\ } This on-going merger galaxy has been the
target of numerous studies. As a member of the Leo triplet, it appears
to have suffered most from the mutual interaction, its stellar disk
being strongly warped. Its activity is manifest over the entire
electromagnetic spectrum, from radio (Reuter et al. \cite{reut91})
through X-rays (Dahlem et al.  \cite{dahl96}). Despite the strongly
elongated synthesised beam of the WSRT at this declination,
significant diffuse radio continuum emission is seen to extend in the
z-direction on either side of the disk, with some bright protusions
marking the footpoints of the most active regions. Strong absorption
in the \ion{H}{i} map marks the location of the active nucleus. OH
absorption has previously been detected in this galaxy by Rickard et
al. \cite{rick82}. This is clearly seen in
Fig.~\ref{fig:hiohabs}. Peak line depths at 1667 and 1665 MHz are
-12.6 and -8.28 mJy/Beam relative to a continuum brightness of 202
mJy/Beam measured with the same PSF. The 1667/1665 line ratio of 1.52
is intermediate between the optically thin and optically thick cases.
The \ion{H}{i} absorption line depth is -18.9 mJy/Beam relative to the
233 mJy/Beam continuum brightness. The 1.3:1 ratio of observed 
\ion{H}{i} to OH opacity is almost as extreme as that detected in
NGC~3079.

\noindent{\it NGC 3938\ } The stellar kinematics of this galaxy was 
studied by Bottema (\cite{bott88}) along with that of NGC~3198. We 
detect a diffuse disk and emission from spiral arms in the radio 
continuum which are also prominent in H$\alpha$ (see the image in 
Jim\'enez-Vicente et al (\cite{jime99}). The \ion{H}{i} traced in 
our observations does not extend much beyond the stellar disk. 

\noindent{\it NGC 4125\ } This is an elliptical galaxy containing
an incipient disk. According to Fabbiano \& Schweizer (\cite{fabi95}) 
its rich fine-structure (they report two plumes or disks crossing 
at right angels) is indicative of a recent merger. We detect a modest 
nuclear source coincident with the stellar galaxy nucleus and clearly 
separated from what appears to be a classical double radio galaxy 
displaced to the South-West. 

\noindent{\it NGC 4236\ } Diffuse disk emission is detected, together 
with knots of enhanced brightness in this member of the M~81 group (like 
NGC~2403, see above). The two brightest radio sources in the field center (a 
double and single source) are likely to be background objects. Despite 
the abundant (atomic) hydrogen, its star formation rate is low, with 
only very diffuse radio emission detectable. 

\noindent{\it NGC 4254\ } This galaxy is at the periphery of the Virgo
Cluster. Its structure and kinematics were recently studied by Vollmer
et al. (\cite{voll05}); its radio polarisation and magnetic field
structure were investigated by Soida et al. (\cite{soid96}). Its
strongly asymmetric structure is evident in both the radio continuum
and the \ion{H}{i}, with faint extensions to the North that go well
beyond the optical disk.

\noindent{\it NGC 4321\ } This classical grand-design spiral, also located 
in the Virgo Cluster, has been the target of numerous studies, e.g. in the 
radio continuum (van der Hulst \cite{vdhu81}), \ion{H}{i} (Knapen et al. 
\cite{knap93}), and in CO (Cepa et al. \cite{cepa92}). The rather 
elongated synthesised beam reduces contrast in our radio continuum
map. The \ion{H}{i} emission is again (as in other Virgo spirals)
truncated near the edge of the optical disk.

\noindent{\it NGC 4450\ } Another member of the Virgo Cluster. Nuclear 
emission and a disk with small scale-length are detected in the radio 
continuum map of this LINER galaxy. It exhibits 
little \ion{H}{i}. 

\noindent{\it NGC 4552\ } This is an elliptical galaxy belonging to
``Sub-cluster~A'', located East of the Virgo Cluster (see Machacek et
al. \cite{mach06}. Its low radial velocity implies that it is moving
supersonically through the Virgo Cluster, thus making it a prime
candidate for (former) ram-pressure stripping. Our radio continuum map
is the first obtained for this galaxy and shows nuclear emission and a
double lobe along PA $\sim$ 80\degr. Such a double lobe is suggestive
of previous activity from what is now a low power AGN in the centre of
this galaxy (Cappellari et al. \cite{capp99}).

\noindent{\it NGC 4559\ } Diffuse disk emisison is detected in this
isolated spiral galaxy. Barbieri et al. (\cite{barb05}) have recently
investigated its structure and kinematics in the \ion{H}{i} line.

\noindent{\it NGC 4569\ } A bright radio continuum disk with a small
scale-length is detected co-aligned with the stellar component. The
most salient feature is a double lobe extension along the minor axis
at PA $\sim$ 100\degr. There are but very few spiral galaxies of this
kind, namely 0421+0400 (Beichman et al. \cite{beic85}), 0313-192 (Keel
et al. \cite{keel02}), NGC~2992 (Wehrle \& Morris \cite{wehr88}),
NGC~3079 (s.a.), NGC~3367 (Garc\'ia-Barreto et al. \cite{garc98}), and
NGC~4258 (e.g. van Albada \& van der Hulst \cite{alba82}). They all
share the common property of being radio-overluminous in the radio-FIR
correlation.

\noindent{\it NGC 4631\ } This edge-on galaxy is well known for its
radio halo, which is also prominent in our radio continuum
map. Numerous filaments extending from the disk into the halo are
apparent. The \ion{H}{i} map shows tidal features caused by the strong
interaction with the nearby galaxy NGC~4656 (see the work of Rand
\cite{rand94}).

\noindent{\it NGC 4725\ } A low surface brightness ring and inner disk
are detected in this galaxy. The ring is coincident with the
star-forming ring for which this galaxy is known (Buta \cite{buta88}).
Its interaction with nearby NGC~4747 (Wevers et al. \cite{weve84}) may
be responsible for the asymmetries seen in our images.

\noindent{\it NGC 4736\ } The kinematics and star formation activity
in the central bar and ring of this LINER galaxy have been the focus
of numerous investigations (e.g. Mu\~noz-Tu\~n\'on et
al. (\cite{muno04}). We detect a high-surface brightness nuclear
region, the inner ring and diffuse outer structures. Note the large
(7\arcmin) background FR-II radio galaxy to the South-West.

\noindent{\it NGC 4826\ } Also known as the ``Black Eye'' or ``Evil
Eye galaxy'', this LINER exhibits a counter-rotating gaseous disk in
its central region (see e.g. Braun et al. \cite{brau94};
Garc\'ia-Burillo et al.  \cite{garc03}). The high column density
\ion{H}{i} in the central regions rotates in one sense, while the
diffuse outer \ion{H}{i} disk in the other. Our radio continuum image
shows a high-surface brightness inner disk in this system.

\noindent{\it NGC 5033\ } This tidally distorted Seyfert-I galaxy has
seen numerous kinematic studies (e.g. Thean et al. \cite{thea97}).
Our \ion{H}{i} map shows substantial asymmetry, the gaseous disk being
strongly warped. The radio continuum image reveals a bright inner- and
faint outer disk.

\noindent{\it NGC 5055\ } This flocculent spiral galaxy possesses a
huge \ion{H}{i} disk (of $\sim$30 arcmin diameter), which is strongly
warped at large radii (Battaglia et al. \cite{batta06}). The diffuse
outer disk is not accurately represented in our \ion{H}{i} image and
the reader is referred to Battaglia et al. (\cite{batta06}) for a better
representation. The radio continuum closely traces the stellar disk
and its spiral structure. Two diffuse arcs are apparent to the
North-west and South-east of the high brightness disk. 

\noindent{\it NGC 5194\ } Easily recognisable as ``M~51'' in any
spectral regime, this prototypical grand-design spiral is one of the
most thoroughly studied. Qualitatively, the radio continuum image is
very similar to the optical (excepting the tidal companion NGC~5195),
the high brightness regions are co-located with the regions of high star
formation rate.

\noindent{\it NGC 6946\ } As was the case for M~51 above, the radio 
continuum readily discloses the identity of this spiral, which 
has also been the target of numerous studies in all spectral bands. 
The \ion{H}{i} extends far beyond the stellar disk, maintaining 
spiral structure as far out as it can be traced. 

\noindent{\it NGC 7331\ } A high-surface brightness inner disk is 
superposed on more diffuse outer arms and disk emission. Its high 
star-forming activity may be connected with the counter-rotation 
of the bulge discovered by Prada et al. (\cite{prad96}). The 
structure of the galaxy is well illustrated in the recent work 
of Regan et al. (\cite{rega04}). The diffuse radio continuum in the
South-East corner of the field is one lobe of a giant (20\arcmin)
head-tail background radio galaxy.

\subsection{Final Comments}
\label{subsec:final}

\begin{figure*}
   \centering
   \includegraphics[width=16.4cm]{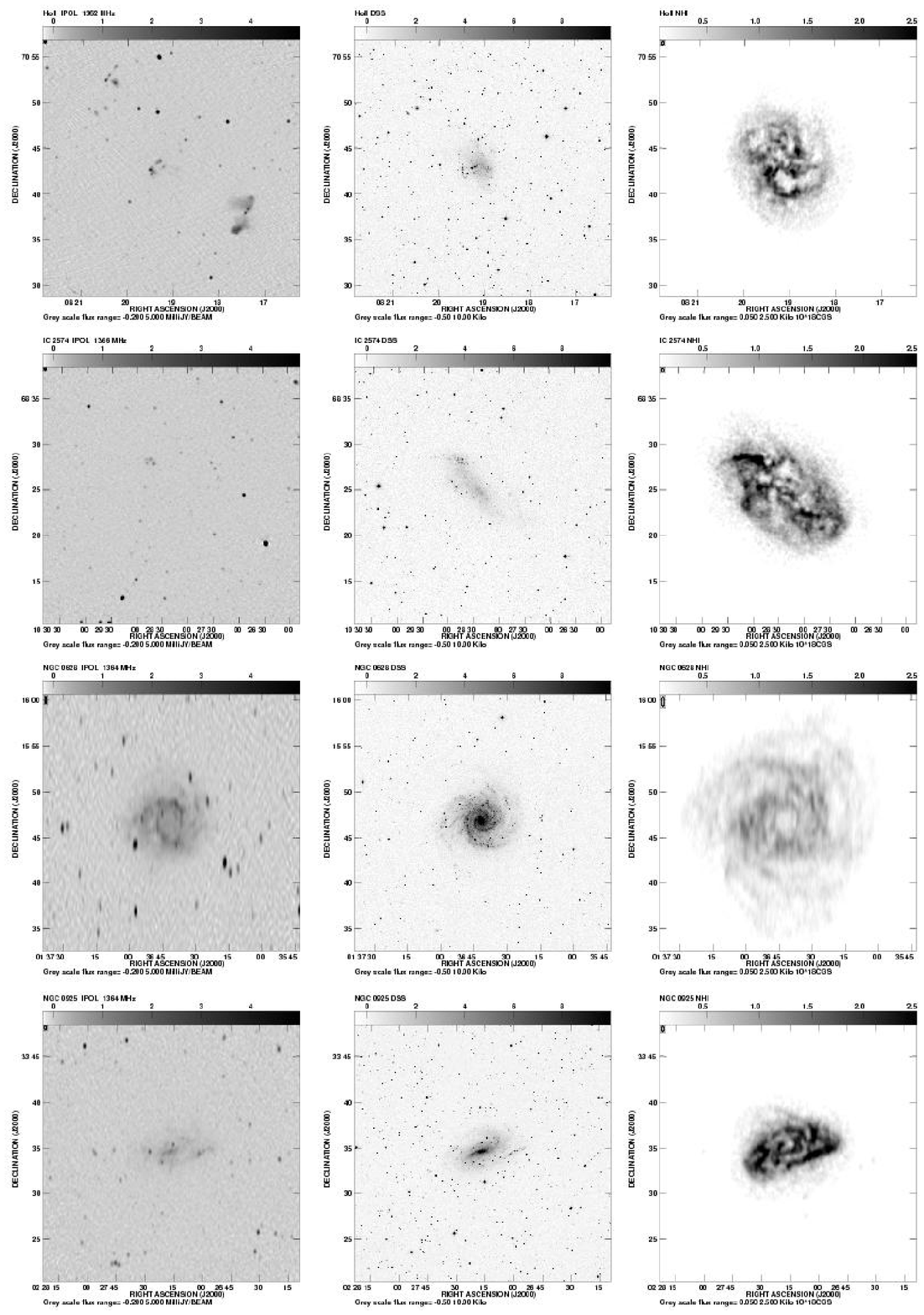}
      \caption{Radio continuum (left), optical POSS-II Red (center)
              and neutral hydrogen column density (right) for the four
              indicated galaxies. A square root transfer
              function is used for the radio continuum image and a
              linear one for the optical (in densitometer counts) and
              HI images.  }
         \label{fig:fourset1}
   \end{figure*}

\begin{figure*}
   \centering
   \includegraphics[width=16.4cm]{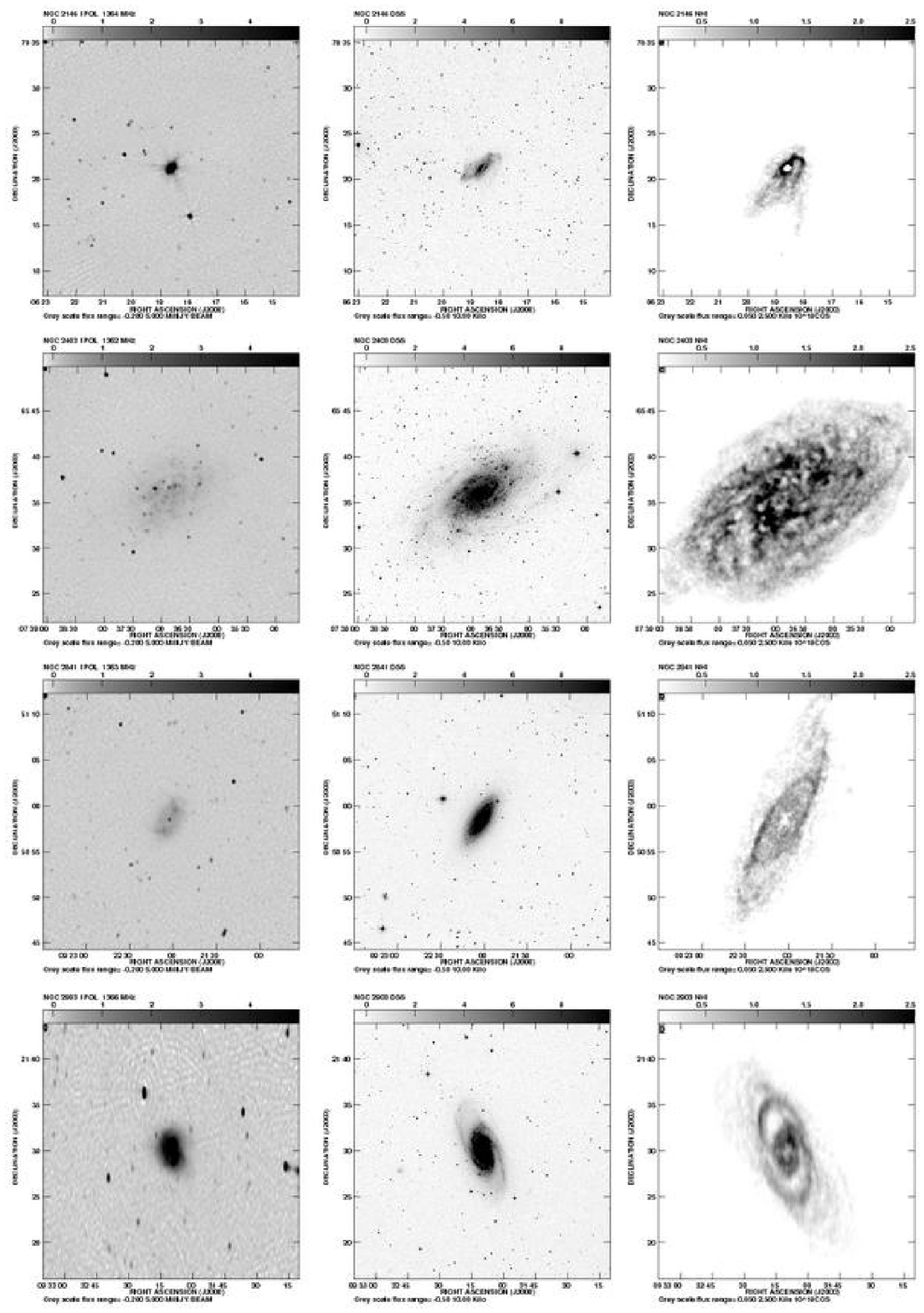}
      \caption{Radio continuum (left), optical POSS-II Red (center)
              and neutral hydrogen column density (right) for the four
              indicated galaxies. A square root transfer
              function is used for the radio continuum image and a
              linear one for the optical (in densitometer counts) and
              HI images.  }
         \label{fig:fourset2}
   \end{figure*}

\begin{figure*}
   \centering
   \includegraphics[width=16.4cm]{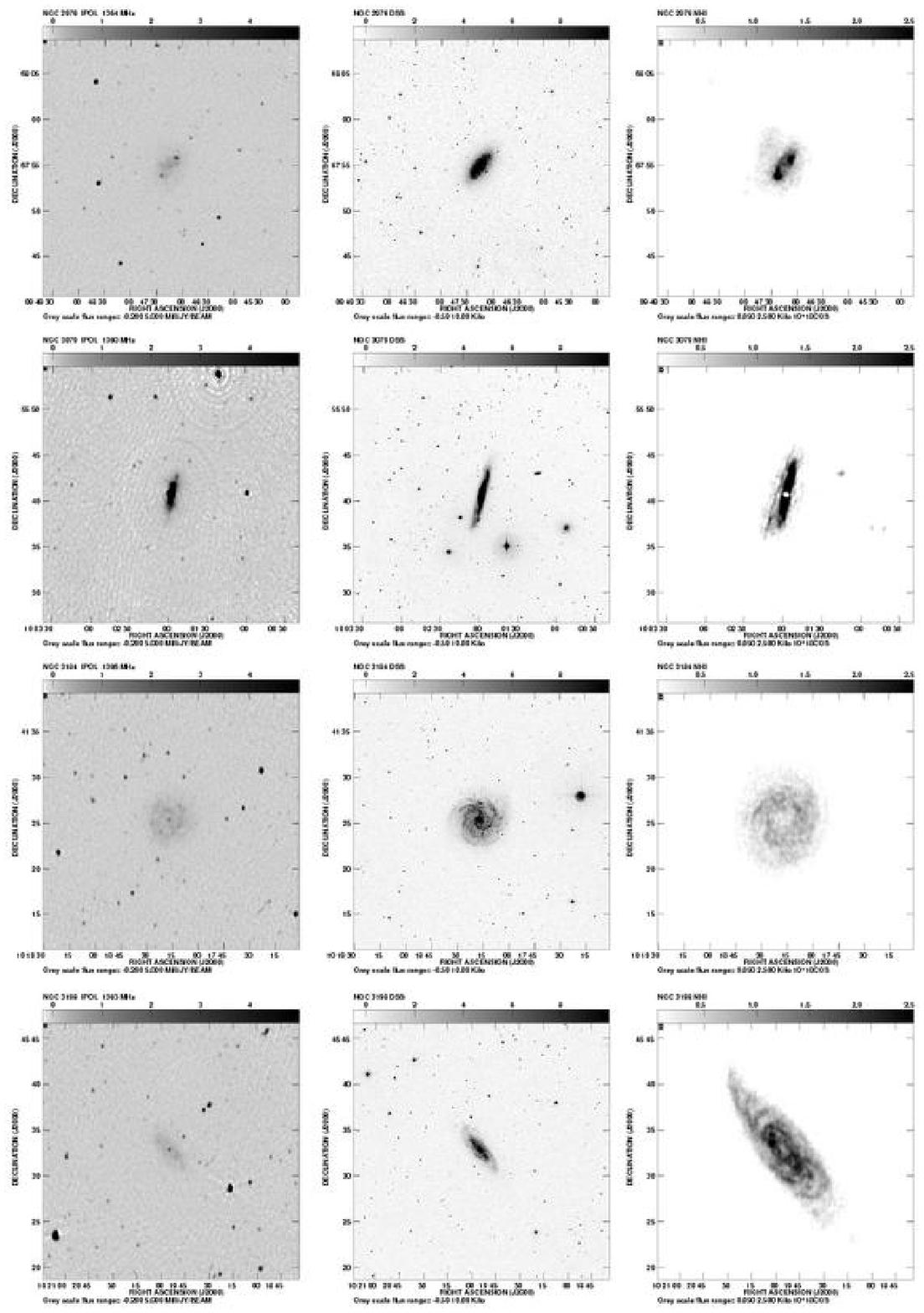}
      \caption{Radio continuum (left), optical POSS-II Red (center)
              and neutral hydrogen column density (right) for the four
              indicated galaxies. A square root transfer
              function is used for the radio continuum image and a
              linear one for the optical (in densitometer counts) and
              HI images.  }
         \label{fig:fourset3}
   \end{figure*}

\begin{figure*}
   \centering
   \includegraphics[width=16.4cm]{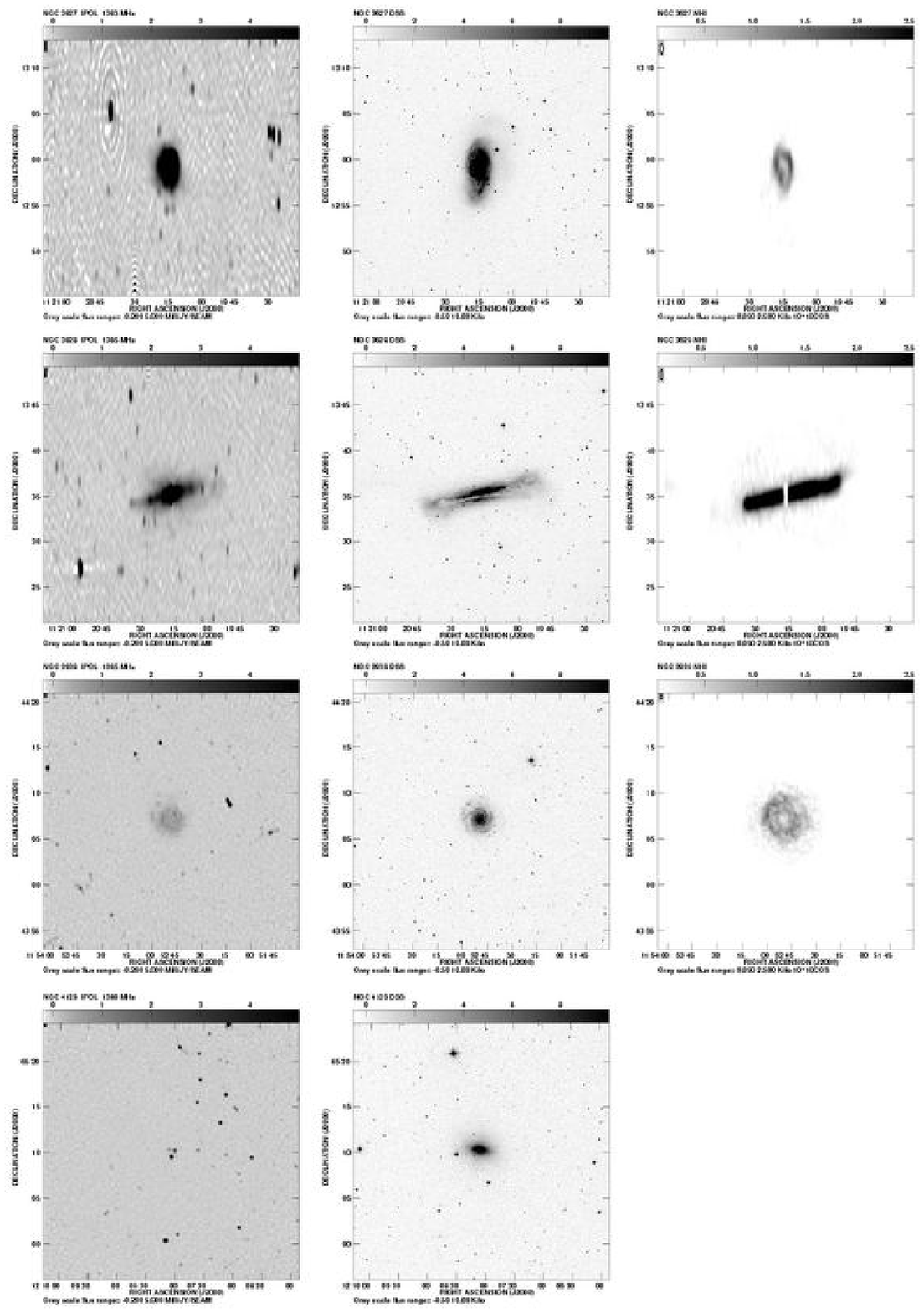}
      \caption{Radio continuum (left), optical POSS-II Red (center)
              and neutral hydrogen column density (right) for the four
              indicated galaxies. A square root transfer
              function is used for the radio continuum image and a
              linear one for the optical (in densitometer counts) and
              HI images.  }
         \label{fig:fourset4}
   \end{figure*}

\begin{figure*}
   \centering
   \includegraphics[width=16.4cm]{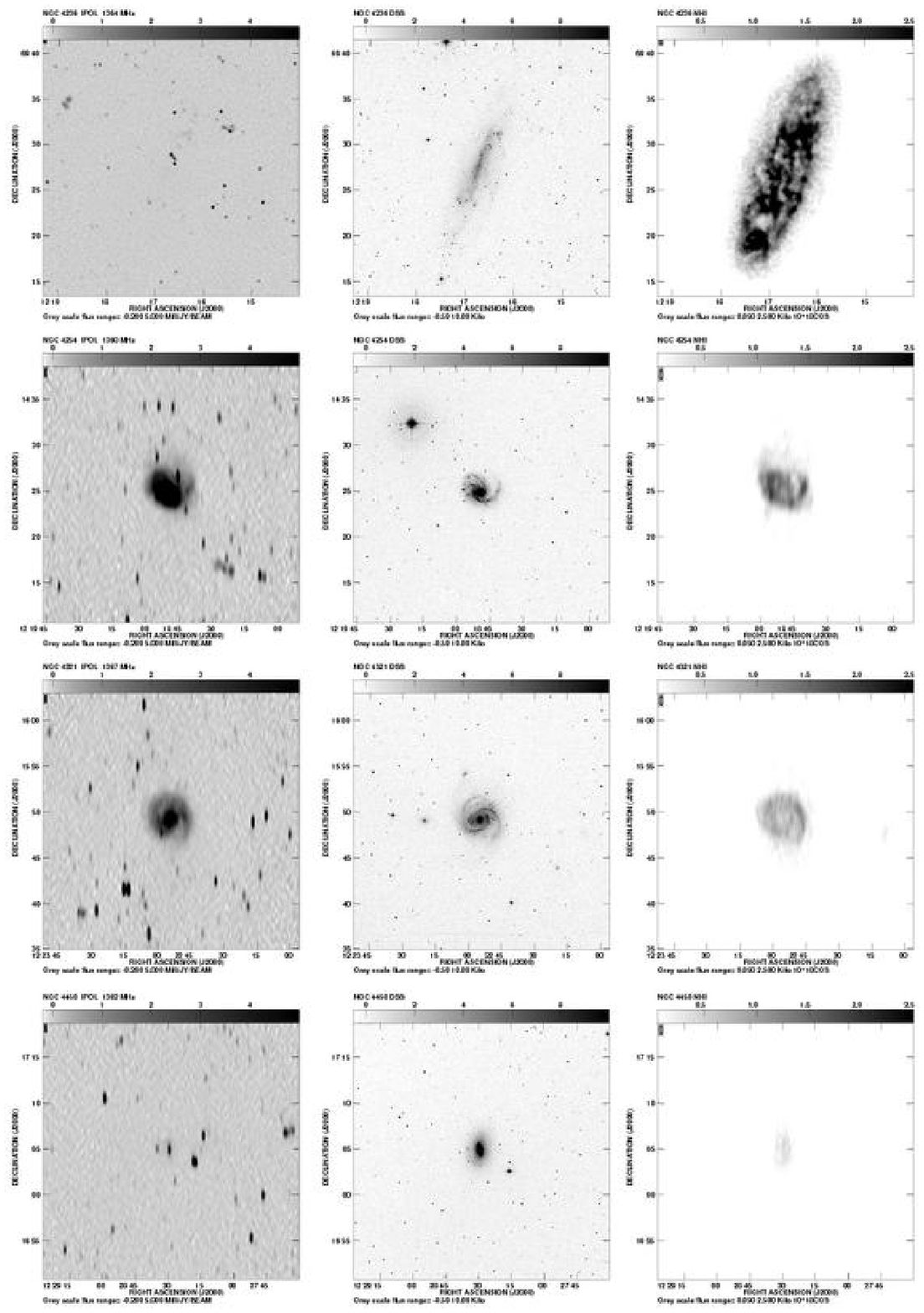}
      \caption{Radio continuum (left), optical POSS-II Red (center)
              and neutral hydrogen column density (right) for the four
              indicated galaxies. A square root transfer
              function is used for the radio continuum image and a
              linear one for the optical (in densitometer counts) and
              HI images.  }
         \label{fig:fourset5}
   \end{figure*}

\begin{figure*}
   \centering
   \includegraphics[width=16.4cm]{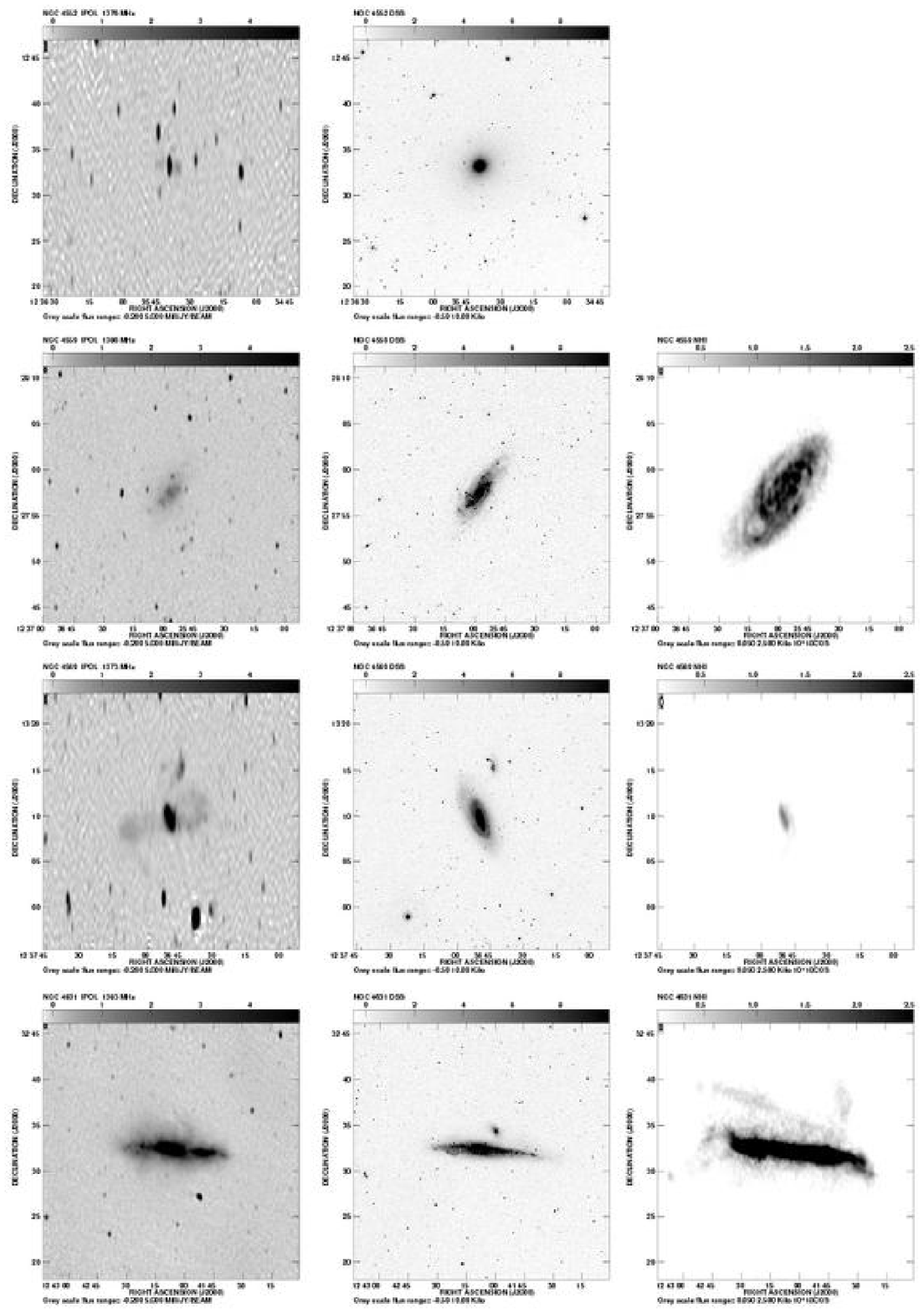}
      \caption{Radio continuum (left), optical POSS-II Red (center)
              and neutral hydrogen column density (right) for the four
              indicated galaxies. A square root transfer
              function is used for the radio continuum image and a
              linear one for the optical (in densitometer counts) and
              HI images.  }
         \label{fig:fourset6}
   \end{figure*}

\begin{figure*}
   \centering
   \includegraphics[width=16.4cm]{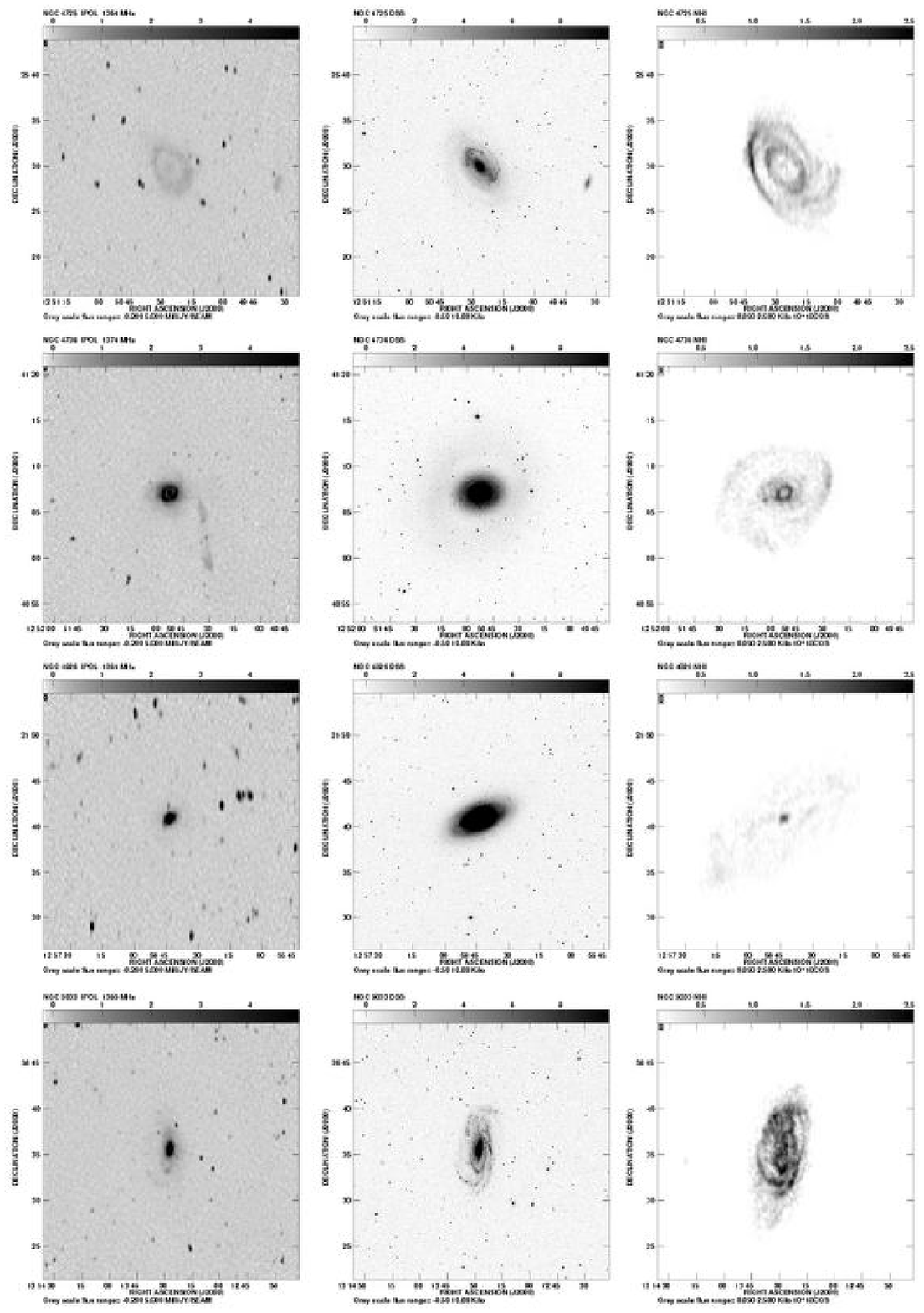}
      \caption{Radio continuum (left), optical POSS-II Red (center)
              and neutral hydrogen column density (right) for the four
              indicated galaxies. A square root transfer
              function is used for the radio continuum image and a
              linear one for the optical (in densitometer counts) and
              HI images.  }
         \label{fig:fourset7}
   \end{figure*}

\begin{figure*}
   \centering
   \includegraphics[width=16.4cm]{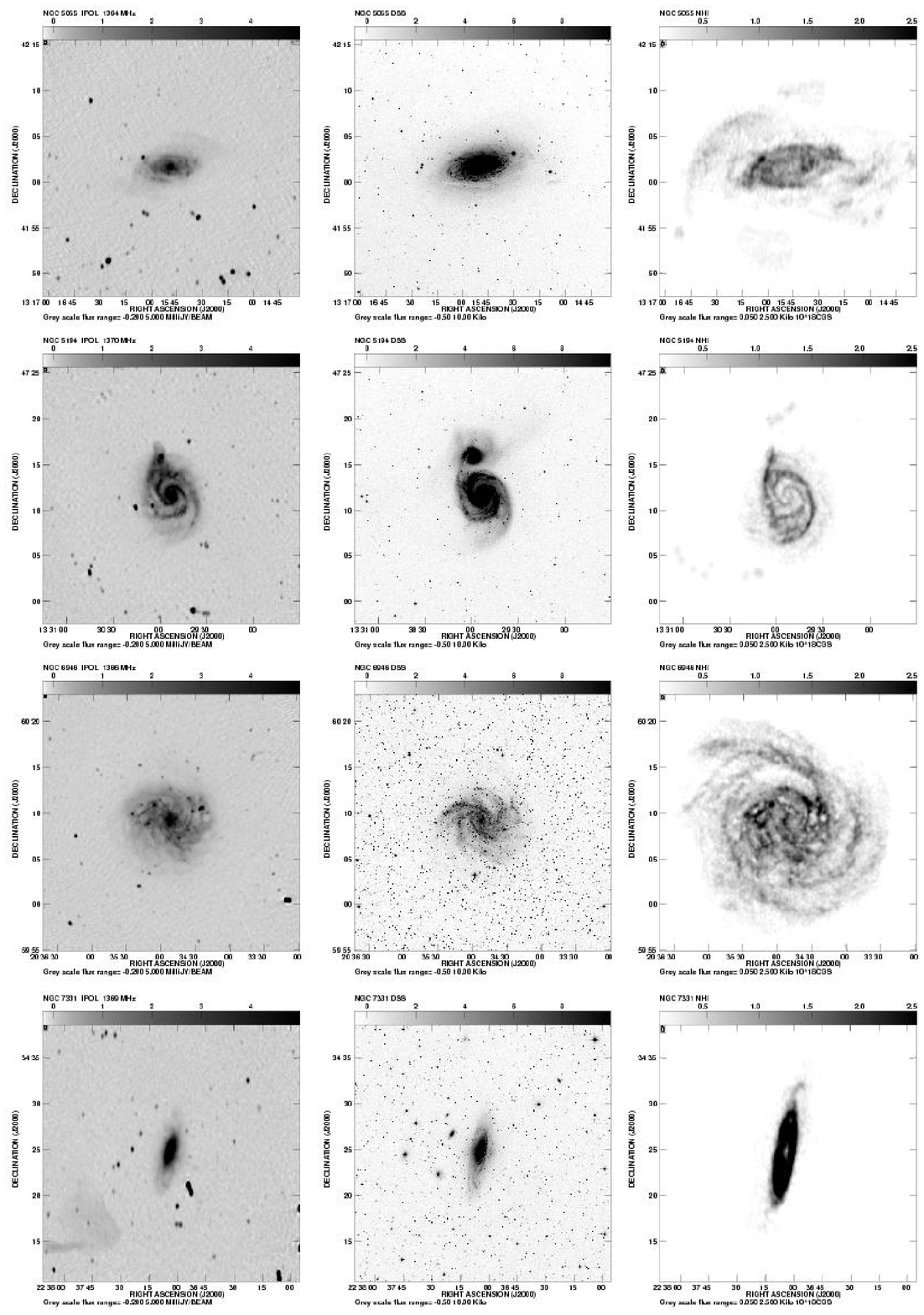}
      \caption{Radio continuum (left), optical POSS-II Red (center)
              and neutral hydrogen column density (right) for the four
              indicated galaxies. A square root transfer
              function is used for the radio continuum image and a
              linear one for the optical (in densitometer counts) and
              HI images.  }
         \label{fig:fourset8}
   \end{figure*}

\begin{figure*}
   \centering
   \includegraphics[width=12cm]{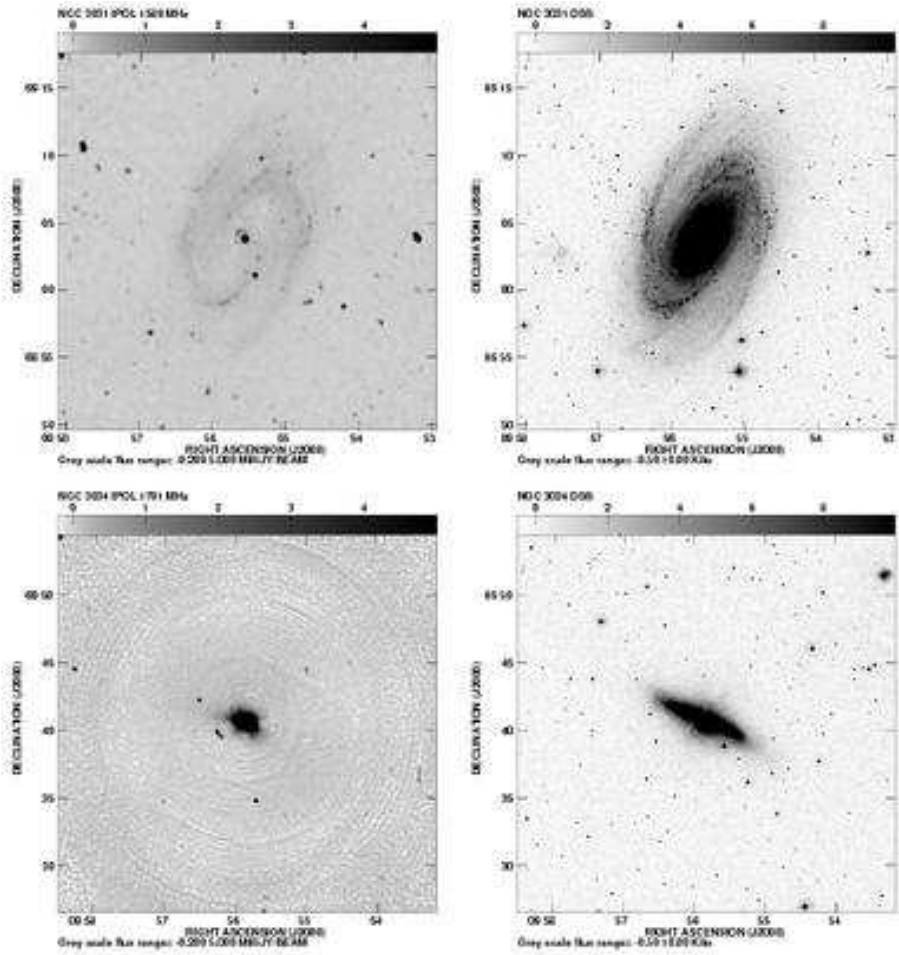}
      \caption{Radio continuum (left) and optical POSS-II Red (right)
              images for the 
              indicated galaxies. A square root transfer
              function is used for the radio continuum image and a
              linear one for the optical (in densitometer counts) image.  }
         \label{fig:twoset}
   \end{figure*}

\begin{figure*}
   \centering
   \includegraphics[width=18cm]{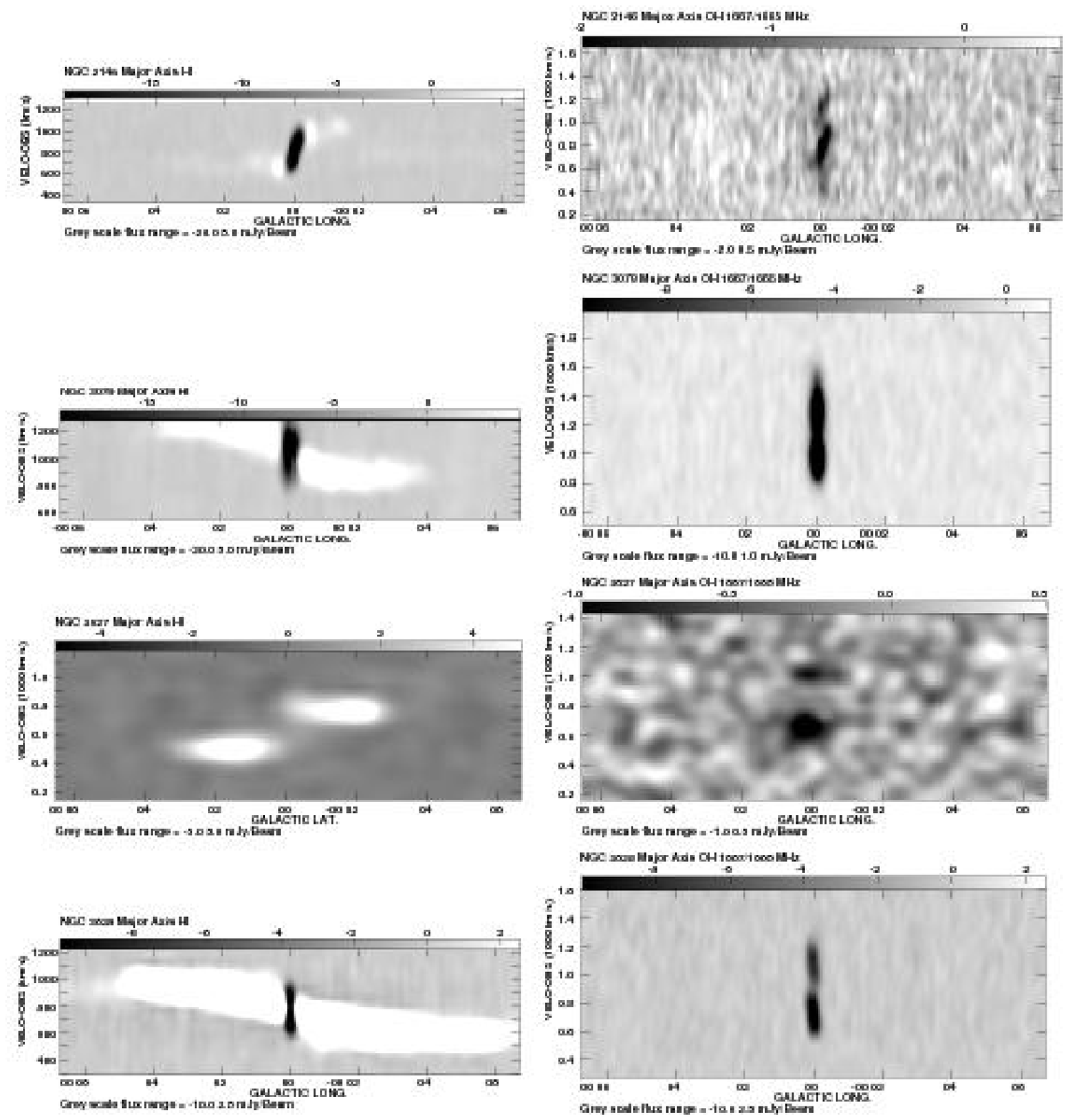}
      \caption{HI emission and absorption (left) and OH 1667/1665 MHz
              absorption (right) along the major axis of the indicated
              galaxies. A linear transfer function is used with the
              indicated range.  }
         \label{fig:hiohabs}
   \end{figure*}

The most useful contribution of the data presented in this paper to the
literature on nearby galaxies is to place their radio continuum emission in a
more general context. Our survey has allowed resolved detection and imaging of
every one of our sub-sample galaxies. The standardized presentation of
Figs.~\ref{fig:fourset1} -- \ref{fig:twoset} allows direct comparison of
galaxies which span a remarkable range of morphological type and star
formation rate; much as the Hubble Atlas (Sandage \cite{sand84}) has allowed
generations of astronomers to appreciate the diverse optical continuum
appearance of nearby galaxies. A similar rich diversity of radio continuum
(and for that matter \ion{H}{i}) appearances becomes obvious in the figures.

Although in this paper we have focused on the presentation of the main
products (radio continuum and \ion{H}{i} total intensity images)
derived from the WSRT observations, the data have the potential to
allow more follow-up and to help address some key questions:

\begin{itemize}
\item 
{Star formation in a wide range of nearby galaxy environments}: 
Radio continuum emission is closely tied to massive star formation
(and death) or nuclear activity. It traces both the prompt (5 Myr)
thermal emission from \ion{H}{ii} regions as well as the longer-lived
($\sim$30 Myr) synchrotron emission of relativistic electrons in the
local magnetic field. Previous studies have shown that at $\sim$1.5~GHz radio
frequencies the non-thermal:thermal ratio of galaxy disk emission is
about 4:1 (eg. Condon \cite{cond92}), while the AGN contribution in
most of our sample galaxies is seen to be minor. Hence the images are
likely to be dominated by the non-thermal continuum emission of
several GeV cosmic ray electrons. 

As shown in Murphy et al. (\cite{murp06}) the {\it Spitzer} Infrared
Nearby Galaxies Survey (SINGS; Kennicutt et al. \cite{kenn03}) and the
WSRT-SINGS radio continuum survey presented here, can be used to study
the effects of star-formation activity on the far-infrared
(FIR)--radio correlation {\it within} galaxies. 
Future modeling of the physical mechanisms underlying the FIR-radio
continuum correlation will certainly benefit from taking explicit
account of the distinct timescales of all of the different emission
mechanisms. These and many other topics will be addressed in
subsequent papers utilizing the multi-spectral SINGS data.

\item{Galactic Magnetic Fields}: a key component of the ISM is the
magnetic field but in order to obtain as realistic picture as possible
of the morphology and strength of the magnetic field in galxies, it is
mandatory to map the nonthermal emission in such a way that a rotation
measure (RM) analysis is rendered feasible. By using all of the
individual frequency channels in each in the eight available sub-bands
it is possible to trace continuous variations of the RM (see for a
complete discussion Brentjens \& de Bruyn \cite{bren05}).

The polarization characteristics of the galaxies in the sample, will be
presented in a forthcoming paper.

\item{\ion{H}{i} and OH emission/absorption}: it is important to
stress again the power of a wide-band, high resolution spectral line
correlator such as currently available at the WSRT.  In this way, we
have been able to extract both the HI content and the presence (or
absence) of OH absorption, in addition to the radio continuum
morphology which was the main goal of the observations. Although not
yet analyzed for the presented dataset, this has also enabled an
unbiased survey of red-shifted emission line objects which fall in
the band (as shown in Morganti et al. \cite{morg04}).

\end {itemize}

\begin{table*}
\caption{Galaxy Observations and Data Attributes}             
\label{tab:obse}      
\centering          
\begin{tabular}{l c c c c c c c }     
\hline\hline       
Name & Observation Dates & 18cm Beam & 18cm {\sc RMS} & 22cm Beam &
22cm {\sc RMS} & HI Beam & HI {\sc RMS}\\
\ &\ &NS$\times$EW ('')&($\mu$Jy/Beam)&NS$\times$EW
('')&($\mu$Jy/Beam)&NS$\times$EW('')& ($\mu$Jy/Beam@625 kHz)\\ 
\hline           
Holmberg II & 2003/03/07 & 10.5$\times$10.0 & 38 & 13.0$\times$12.5 & 27&24.8x17.6&140\\
IC 2574 & 2003/03/23 & 10.5$\times$10.0 & 36 & 13.0$\times$12.5 & 24&20.1x19.1&120\\
NGC 0628 & 2003/03/09 & 36.5$\times$10.0 & 40 & 45.5$\times$12.5 & 29&62.5x16.9&160\\
NGC 0925 & 2003/03/22 & 18.0$\times$10.0 & 36 & 22.5$\times$12.5 & 26&31.7x19.2&140\\
NGC 2146 & 2003/03/08 & 10.0$\times$10.0 & 40 & 12.5$\times$12.5 & 29&19.0x18.4&150\\
NGC 2403 & 2003/04/06 & 10.5$\times$10.0 & 37 & 13.5$\times$12.5 & 26&20.9x19.6&140\\
NGC 2841 & 2003/03/10 & 12.5$\times$10.0 & 35 & 16.0$\times$12.5 & 25&24.0x20.1&130\\
NGC 2903 & 2003/03/26 & 27.0$\times$10.0 & 38 & 34.0$\times$12.5 & 32&45.4x18.3&140\\
NGC 2976 & 2003/03/21 & 10.5$\times$10.0 & 36 & 13.0$\times$12.5 & 24&20.9x19.7&140\\
NGC 3031 & 2003/04/17+05/05 & -- & -- & 14.5$\times$13.6 & 24 & -- & --\\
NGC 3034 & 2003/03/16 &10.5$\times$10.0 & 75 & -- & -- & -- & --\\ 
NGC 3079 & 2003/03/27 & 12.0$\times$10.0 & 40 & 15.0$\times$12.5 & 34&23.1x21.1&170\\
NGC 3184 & 2003/05/10 & 15.0$\times$10.0 & 36 & 18.5$\times$12.5 & 25&27.2x19.2&140\\
NGC 3198 & 2003/04/09 & 14.0$\times$10.0 & 37 & 17.5$\times$12.5 & 27&25.6x20.2&150\\
NGC 3627 & 2003/04/10 & 44.0$\times$10.0 & 46 & 55.5$\times$12.5 & 42&75.5x17.3&170\\
NGC 3628 & 2003/04/08 & 42.5$\times$10.0 & 43 & 53.0$\times$12.5 & 32&72.0x17.0&160\\
NGC 3938 & 2003/03/13 & 14.0$\times$10.0 & 36 & 17.5$\times$12.5 & 26&25.2x19.3&140\\
NGC 4125 & 2003/03/17 & 11.0$\times$10.0 & 107 & 13.5$\times$12.5 & 27&24.1x22.1&150\\
NGC 4236 & 2003/05/08 & 10.5$\times$10.0 & 42 & 13.0$\times$12.5 & 24&20.6x19.7&140\\
NGC 4254 & 2003/04/15 & 40.0$\times$10.0 & 44 & 50.0$\times$12.5 & 34&68.6x17.0&160\\
NGC 4321 & 2003/04/16 & 36.5$\times$10.0 & 42 & 45.5$\times$12.5 & 31&62.6x17.2&180\\
NGC 4450 & 2003/04/21 & 34.0$\times$10.0 & 39 & 42.5$\times$12.5 & 27&58.1x17.4&170\\
NGC 4552 & 2003/07/10 & 46.0$\times$10.0 & 90 & 57.5$\times$12.5 & 35&77.6x17.2&160\\
NGC 4559 & 2003/06/30 & 21.0$\times$10.0 & 53 & 26.5$\times$12.5 & 26&36.3x19.2&150\\
NGC 4569 & 2003/03/24 & 43.5$\times$10.0 & 73 & 54.5$\times$12.5 & 26&74.1x17.3&150\\
NGC 4631 & 2003/04/03 & 18.5$\times$10.0 & 40 & 23.0$\times$12.5 & 27&32.1x18.9&150\\
NGC 4725 & 2003/07/04 & 23.0$\times$10.0 & 43 & 29.0$\times$12.5 & 27&39.1x18.3&180\\
NGC 4736 & 2003/07/06 & 15.0$\times$10.0 & 56 & 19.0$\times$12.5 & 32&27.5x19.6&140\\
NGC 4826 & 2003/06/21 & 27.0$\times$10.0 & 142 & 33.5$\times$12.5 & 28&45.1x18.0&150\\
NGC 5033 & 2003/04/01 & 16.5$\times$10.0 & 38 & 20.5$\times$12.5 & 29&30.1x19.1&150\\
NGC 5055 & 2003/05/06 & 14.5$\times$10.0 & 39 & 18.5$\times$12.5 & 27&30.6x16.6&150\\
NGC 5194 & 2003/03/25+11/23 & 13.5$\times$10.0 & 28 & 17.0$\times$12.5 & 22&24.1x19.3&100\\
NGC 6946 & 2003/03/23+11/22 & 11.5$\times$10.0 & 28 & 14.0$\times$12.5 & 20&21.5x19.3&100\\
NGC 7331 & 2003/03/30+07/09 & 17.5$\times$10.0 & 31 & 22.0$\times$12.5 & 23&33.1x17.9&110\\
\hline                  
\end{tabular}
\end{table*}

\begin{table*}
\caption{Galaxy Properties}             
\label{tab:prop}      
\centering          
\begin{tabular}{l c c c c c }     
\hline\hline       
Name & 1365 MHz Flux $\pm$ Error & Aperture Radius & Method & HI Flux
$\pm$ Error& Peak N$_{HI}$\\
\ & (mJy) & (arcsec) &\ &(Jy-km/s)& 10$^{21}$cm$^{-2}$\\

\hline           
Holmberg II & 5.5 $\pm$ 2 & 65 & Poly & 246 $\pm$ 10&3.08\\
IC 2574 & 14 $\pm$ 2 & 200 & Poly & 413 $\pm$ 20&3.89\\
NGC 0628 & 200 $\pm$ 10 & 450 & Rad & 381 $\pm$ 15&1.44\\
NGC 0925 & 90 $\pm$ 10 & 450 & Blk/Rad & 272 $\pm$ 12&3.58\\
NGC 2146 & 1100 $\pm$ 10 & 200 & Blk/Rad & 72.4 $\pm$ 3&6.33\\
NGC 2403 & 360 $\pm$ 30 & 800 & Blk/Rad & 1030: &4.47\\
NGC 2841 & 100 $\pm$ 7 & 400 & Blk/Rad & 195 $\pm$ 10&1.89\\
NGC 2903 & 460 $\pm$ 10 & 300 & Blk/Rad & 232 $\pm$ 10&2.47\\
NGC 2976 & 68 $\pm$ 5 & 250 & Blk/Rad & 57.3 $\pm$ 3&3.85\\
NGC 3079 & 890 $\pm$ 10 & 300 & Blk/Rad & 102 $\pm$ 5&8.13\\
NGC 3184 & 80 $\pm$ 5 & 300 & Blk/Rad & 112 $\pm$ 5&1.47\\
NGC 3198 & 49 $\pm$ 5 & 300 & Blk/Rad & 220 $\pm$ 10&2.65\\
NGC 3627 & 500 $\pm$ 10 & 350 & Blk/Rad & 35.2 $\pm$ 2&1.85\\
NGC 3628 & 590 $\pm$ 10 & 400 & Blk/Rad & 240 $\pm$ 10&5.15\\
NGC 3938 & 80 $\pm$  5 & 250 & Blk/Rad & 69.8 $\pm$ 4&1.64\\
NGC 4125 & 1.9 $\pm$ 0.2 & 25 & GFit & $<$ 1&$<$ 0.1\\
NGC 4236 & 26 $\pm$ 5 & 600 & Poly & 619 $\pm$ 30&4.74\\
NGC 4254 & 510 $\pm$ 10 & 300 & Rad & 83.5 $\pm$ 4&2.01\\
NGC 4321 & 310 $\pm$ 10 & 250 & Rad & 57.2 $\pm$ 3&1.23\\
NGC 4450 & 13 $\pm$ 2 & 100 & Box & 5.87 $\pm$ 1&0.44\\
NGC 4552 & 93 $\pm$ 5 & 120 & Box & $<$ 1&$<$ 0.1\\
NGC 4559 & 110 $\pm$ 10 & 450 & Rad & 306 $\pm$ 15&3.15\\
NGC 4569 & 170 $\pm$ 10 & 450 & Blk/Rad & 8.52 $\pm$ 1&1.33\\
NGC 4631 & 1290 $\pm$ 10 & 450 & Rad & 511 $\pm$ 25&10.97\\
NGC 4725 & 100 $\pm$ 10 & 600 & Blk/Rad & 136 $\pm$ 6&2.42\\
NGC 4736 & 320 $\pm$ 10 & 400 & Blk/Rad & 85.7 $\pm$ 4&2.00\\
NGC 4826 & 110 $\pm$ 10 & 150 & Rad & 52.1 $\pm$ 2&1.69\\
NGC 5033 & 240 $\pm$ 10 & 450 & Rad & 181 $\pm$ 10&2.78\\
NGC 5055 & 450 $\pm$ 10 & 450 & Rad & 280:&2.59\\
NGC 5194 & 1420 $\pm$ 10 & 450 & Rad & 159 $\pm$ 7&2.19\\
NGC 6946 & 1700 $\pm$ 10 & 600 & Rad & 680 $\pm$ 35&3.14\\
NGC 7331 & 590 $\pm$ 10 & 500 & Blk/Rad & 201 $\pm$ 10&4.59\\
\hline                  
\end{tabular}
\end{table*}

%
%
%
%
%
%
%
%
%
%
%
%
%
%

\begin{acknowledgements}

The Westerbork Synthesis Radio Telescope is operated by ASTRON
(Netherlands Foundation for Research in Astronomy) with support from
the Netherlands Foundation for Scientific Research (NWO). The
Digitized Sky Surveys were produced at the Space Telescope Science
Institute under U.S. Government grant NAG W-2166. The images of these
surveys are based on photographic data obtained using the Oschin
Schmidt Telescope on Palomar Mountain and the UK Schmidt
Telescope. 

\end{acknowledgements}


\begin{thebibliography}{}

   \bibitem[1982]{alba82} van Albada, G.D., van der Hulst, J.M., 
      1982, A\&A, 115, 263 

   \bibitem[1985]{alba85} van Albada, G.D., Bahcall, J.N., Begeman, K., 
      Sancisi, R., 1985, A\&A, 295, 305 

   \bibitem[2006]{batta06} Battaglia, G., Fraternali, F., Oosterloo, T., 
      Sancisi, R., 2006, A\&A, 447, 49 

   \bibitem[1989]{bege89} Begeman, K.G., 1989, A\&A, 223, 47 

   \bibitem[1988]{bott88} Bottema, R., 1988, A\&A, 197, 105 

   \bibitem[1988]{buta88} Buta, R., 
      1988, ApJS, 66, 233

   \bibitem[2005]{barb05} Barbieri, C.V., Fraternali, F., Oosterloo, T.,
      et al., 2005, A\&A, 439, 947 

   \bibitem[1988]{beic85} Beichman, C., Wynn-Williams, C.G., Lonsdale, 
      C.J., et al., 1988, ApJ, 293, 148

   \bibitem[1981]{bosm81} Bosma, A., 1981, AJ, 86, 1791

   \bibitem[1993]{brai93} Braine, J., Combes, F., Casoli, F., et al., 
      1993, A\&AS, 97, 887

   \bibitem[1994]{brau94} Braun, R., Walterbos, R.A.M., Kennicutt, R.C., 
      Tacconi, L., 1994, ApJ, 420, 558 

   \bibitem[2005]{bren05} Brentjens M.A. \& de Bruyn A.G., 2005 A\&A 441, 931

   \bibitem[1991]{brou91} Brouillet, N., Baudry, A., Combes, F., et al., 
      1991, A\&A 242, 35 

   \bibitem[1998]{brou98} Brouillet, N., Kaufman, M., Combes, F., et al., 
      1998, A\&A 333, 92 

   \bibitem[1976]{bruy76} de Bruyn, A. G., Crane, P. C., Price, R. M., 
      Carlson, J. B., 1976, A\&A, 46, 243

   \bibitem[1999]{capp99} Cappellari, M., Renzini, A., Greggio, L., et al., 
      1999, ApJ, 519, 117 

   \bibitem[1992]{cepa92} Cepa, J., Beckman, J.E., Knapen, J., et al., 
      1992, AJ, 103, 429 

   \bibitem[2003]{chem03} Chemin, L., Cayatte, V., Balkowski, C., et al., 
      2003, A\&A, 405, 89 

   \bibitem[1992]{cond92} Condon, J.J., 1992, ARAA, 30, 575

   \bibitem[1996]{dahl96} Dahlem, M., Heckman, T.M., Fabbiano, G.,et al.,
      1996, ApJ, 461, 724 

   \bibitem[2004]{doan04} Doane, N.E., Sanders, W.T., Wilcots, E.M., 
      Juda, M., 1999, AJ, 128, 2712

   \bibitem[1999]{dris99} Drissen, L., Roy, J.-R., Moffat, A.F.J., 
      Shara, M.M., 1999, AJ, 117, 1249 

   \bibitem[1988]{duri88} Duric, N., Seaquist, E.R., 1988,
      ApJ, 326, 574

   \bibitem[1995]{fabi95} Fabbiano, G., Schweizer, F., 1995,
      ApJ, 447, 572

   \bibitem[1988]{fili88} Filippenko, A. V., Sargent, W. L. W., 1988, 
      ApJ, 324, 134 

   \bibitem[2001]{frat01} Fraternali, F., Oosterloo, T., Sancisi, R., 
      van Moorsel, G., 2004, ApJL, 562, L47

   \bibitem[2004]{frat04} Fraternali, F., Oosterloo, T., Sancisi, R., 
      2004, A\&A, 424, 485

   \bibitem[1998]{garc98} Garc\'ia-Barreto, J.A., Rudnick, L., Franco, J., 
      Martos, M., 1998, AJ, 116, 111 

   \bibitem[2003]{garc03} Garc\'ia-Burillo, S., Combes, F., Hunt, L.K., 
      et al., 2003, A\&A, 407, 485 

   \bibitem[1998]{gree98} Greenawalt, B., Walterbos, R.A.M., Thilker,
      D., Hoopes, C.G., 1998, ApJ, 506, 135

   \bibitem[1985]{hasc85} Haschick, A.D., Baan, W.A., 1985, Nature,
   314, 144

   \bibitem[1969]{heil69} Heiles, C., 1969, ApJ, 157, 123

   \bibitem[1981]{vdhu81} van der Hulst, J.M., Crane, P.C., Keel, W.C., 
      1981, AJ, 86, 1175 

   \bibitem[1982]{humm82} Hummel, E., Bosma, A., 1982,
      AJ, 87, 242

   \bibitem[1987]{irwi87} Irwin, J.A., Seaquist, E.R., Taylor, A.R., 
      Duric, N., 1987, ApJL, 313, L91

   \bibitem[1991]{jack91} Jackson, J.M., Eckart, A., Cameron, M., et al., 
      1991, ApJ, 375, 105

   \bibitem[1999]{jime99} Jim\'enez-Vicente, J., Battaner, E., Rozas, M. 
      et al., 1999, A\&A, 342, 417

   \bibitem[1992]{kamp92} Kamphius, J., Briggs, F., 
      1992, A\&A, 253, 335

   \bibitem[2002]{keel02} Keel, W.C., Ledlow, M.J., Owen, F.N., 2002,
      BAAS, Vol. 34, p.1245

   \bibitem[1988]{kenn98} Kennicutt, R. C., 1998,
      ApJ, 498, 541

   \bibitem[2003]{kenn03} Kennicutt, R. C., Armus, L., Bendo, G. et al., 
      2003, PASP, 115, 928

   \bibitem[1993]{knap93} Knapen, J., Cepa, J., Beckman, J.E., et al., 
      1998, ApJ, 416, 563
   
   \bibitem[1996]{lisz96} Liszt, H., Lucas, R., 1996, A\&A, 314, 917

   \bibitem[2006]{mach06} Machacek, M., Jones, W.R., Nulsen, P., 
      2006, ApJ, 644, 155 

   \bibitem[2004]{morg04} Morganti R., Garrett M., Chapman S., et al., 
      2004, A\&A 424, 371 

   \bibitem[2004]{muno04} Mu\~noz-Tu\~n\'on, C., Caon, N., 
      Aguerri, J.A.L., 2004, AJ 127, 58 

   \bibitem[2006]{murp06} Murphy, E.J., Braun, R., Helou, G., et
      al., 2006, ApJ 638, 157
   
   \bibitem[1998]{pisa98} Pisano, D.J., Wilcots, E.M., Elmegreen,
   B.G., 1998, AJ, 115, 975 

   \bibitem[1996]{prad96} Prada, P., Gutierrez, C.M., Peletier, R.F., 
      McKeith, C.D., 1996, ApJL, 463, L5

   \bibitem[1994]{rand94} Rand, R.J., 1994, A\&A, 285, 833

   \bibitem[1999]{rega99} Regan, M.W., Sheth, K., Vogel, S.N., Stuart, N.,  
      1999, ApJ, 526, 97 

   \bibitem[2004]{rega04} Regan, M.W., Thornley, M.D., Bendo, G.J., et al.,  
      2004, ApJS, 154, 204 

   \bibitem[1991]{reut91} Reuter, H.-P., Krause, M., Wielebinski, R., 
      Lesch, H., 1991, A\&A, 248, 12

   \bibitem[1992]{reut92} Reuter, H.-P., Klein, U., Lesch, H., et al., 
      1992, A\&A, 256, 10 

   \bibitem[1982]{rick82} Rickard, L.J., Bania, T.M., Turner, B.E.,
   1982, ApJ 252, 147

   \bibitem[1980]{rots80} Rots, A.H., 1980, A\&AS, 41, 189

   \bibitem[1994]{saik94} Saikia, D.J., Pedlar, A., Unger, S., Axon, D.J., 
      1994, MNRAS, 270, 46

   \bibitem[1984]{sand84} Sandage, A., 1984, {\it Hubble Atlas of Galaxies\
   }, Carnegie Inst. of Washington

   \bibitem[2003]{simo03} Simon, J.D., Bolatto, A.D., Leroy, A., Blitz, L.,  
      2003, ApJ, 596, 957

   \bibitem[1994]{smit94} Smith, B.J., Harvey, P.M., Colome, C., et al.,  
      1994, ApJ, 425, 91

   \bibitem[1996]{soid96} Soida, M., Urbanik, M., Beck, R.,  
      1996, A\&A, 312, 409

   \bibitem[2001]{soid01} Soida, M., Urbanik, M., Beck, R, et al.,  
      2001, A\&A, 378, 40

   \bibitem[2001]{tara01} Taramopoulos, A., Payne, H., Briggs, F.H., 
      2001, A\&A, 365, 360

   \bibitem[2000]{tarc00} Tarchi, A., Neininger, N., Greve, A., et al., 
      2000, A\&A, 358, 95

   \bibitem[1997]{thea97} Thean, A.H.C., Mundel, C.G., Pedlar, A., 
      Nicholson, R.A., 1997, MNRAS, 290, 15

   \bibitem[1995]{tong95} Tongue, T.D., Westphal, D.J., 1995, 
      AJ 109, 2462 

   \bibitem[2000]{voll05} Vollmer, B., Huchtmeier, W., van Driel, W.,  
      2005, A\&A, 439, 92

   \bibitem[1998]{walt98} Walter, F., Kerp, J., Duric, N., Brinks, E.,  
      Klein, U. 1998, ApJ, 502, L147

   \bibitem[1995]{wakk95} Wakker, B.P., Adler, D.S., 1988,
      AJ, 109, 134

   \bibitem[1988]{wehr88} Wehrle, A., Morris, M., 1988, 
      AJ, 95, 1689

   \bibitem[1984]{weve84} Wevers, B.M.H.R., Appleton, P.N., Davies, R.D., 
      Hart, L., 1984, A\&A, 140, 125

   \bibitem[1986]{weve86} Wevers, B.M.H.R., van der Kruit, P.C.,
      Allen, R.J., 1986, A\&AS, 66, 505

   \bibitem[1992]{whit92} White, R.L., Becker, R.H., 1992, ApJS 79, 331

   \bibitem[1993]{zhan93} Zhang, X., Wright, M., Alexander, P., 
      1993, ApJ, 418, 100 
   

\end{thebibliography}
\end{document}